\documentclass[sigconf]{acmart}

\usepackage{makecell}
\usepackage{graphicx}
\usepackage{subfigure}
 
 \usepackage[flushleft]{threeparttable}
 
\usepackage{booktabs}
\usepackage{multirow}
\usepackage{hyperref}
\usepackage{url}
\usepackage{breqn}

\usepackage{xcolor,colortbl}
\definecolor{darkgreen}{rgb}{0.0, 0.2, 0.13}
\definecolor{darkred}{rgb}{0.2, 0.0, 0.13}

\newcommand{\acc}[1]{\cellcolor{darkgreen!30!white!#1}}

\newcommand{\BfPara}[1]{{\noindent\bf#1.}\xspace}
\usepackage{multicol}

\usepackage[skip=5pt]{caption}

\usepackage{slashbox}

\hypersetup{
	pdfpagemode=pagewidth,
	plainpages=false,
	colorlinks,
	urlcolor=blue,
	linkcolor=red,
	citecolor=green,
	bookmarksnumbered
}

\usepackage{xspace}
\usepackage{setspace}

\newcommand{\etal}{{\em et al.}\xspace}

\newcommand{\vs}[1]{{\vspace{-#1mm}}}

\usepackage[inline,shortlabels]{enumitem}

\newcommand*\circled[1]{\tikz[baseline=(char.base)]{%
            \node[shape=circle,fill=black,draw,text=white,inner sep=0.5pt] (char) {#1};}}
\usepackage{tikz}

\begin{document}


\title[Cleaning the NVD]{Cleaning the NVD: Comprehensive Quality Assessment, Improvements, and Analyses}

\author{Afsah Anwar}
\affiliation{\institution{University of Central Florida}}
\email{afsahanwar@knights.ucf.edu} 

\author{Ahmed Abusnaina}
\affiliation{\institution{University of Central Florida}}
\email{ahmed.abusnaina@knights.ucf.edu}

\author{Songqing Chen}
\affiliation{\institution{George Mason University}}
\email{sqchen@gmu.edu}

\author{Frank Li}
\affiliation{\institution{Georgia Institute of Technology}}
\email{frankli@gatech.edu}

\author{David Mohaisen}
\affiliation{\institution{University of Central Florida}}
\email{mohaisen@cs.ucf.edu}


\begin{abstract}
Vulnerability databases are vital sources of information on emergent software security concerns. Security professionals, from system administrators to developers to researchers, heavily depend on these databases to track vulnerabilities and analyze security trends. How reliable and accurate are these databases though?

In this paper, we explore this question with the National Vulnerability Database (NVD), the U.S. government's repository of vulnerability information that arguably serves as the industry standard.
Through a systematic investigation, we uncover inconsistent or incomplete data in the NVD that can impact its practical uses, affecting information such as the vulnerability publication dates, names of vendors and products affected, vulnerability severity scores, and vulnerability type categorizations. We explore the extent of these discrepancies and identify methods for automated corrections. Finally, we demonstrate the impact that these data issues can pose by comparing analyses using the original and our rectified versions of the NVD. Ultimately, our investigation of the NVD not only produces an improved source of vulnerability information, but also provides important insights and guidance for the security community on the curation and use of such data sources.
\end{abstract}



\maketitle

\section{Introduction}\label{sec:introduction}



Securing computer systems in practice entails identifying, understanding, and remediating the stream of software security concerns that are continuously uncovered. To effectively do so, security professionals and researchers depend on various sources of information to inform them of new security issues. One vital source is vulnerability databases, which operate as a repository of vulnerability information. However, is the information actually reliable?

In this work, we explore this question by identifying the limitations of existing vulnerability datasets and their implications on real-world security operations.
While several vulnerability databases exist, we focus on the one that is arguably the most widely used: the National Vulnerability Database (NVD). This database, maintained by the US government, strives to accurately document all publicly known vulnerabilities, and effectively serves as the industry's standard. Both commercial security services (e.g., Hakiri~\cite{hakiri}, Snyk~\cite{synk}, and SourceClear~\cite{sourceclear}), and open-source security tools (e.g., Bundler-audit~\cite{bundaud}, OWASP OSSIndex~\cite{ossindex}, and Dependency-check~\cite{depcheck}) depend on the NVD's vulnerability information to function effectively.
Furthermore, researchers~\cite{erdHosicommon,AnwarKNM2018,LiP17} have used the NVD as a core data source to shed light on aspects of the vulnerability discovery and remediation process.
Given the importance of the NVD, it is crucial that we understand the quality of its data, lest some incorrect information leads to a critical security lapse~\cite{wannacry}. 

The prior work~\cite{MuCYHXMW18,DongGCX19,NguyenM13,LiP17} has investigated certain types of data quality concerns in NVD. However, to the best of our knowledge, there has not been a systematic and comprehensive analysis of inconsistencies and incomplete data in the NVD to date. To close this gap, in this paper, we perform an in-depth large-scale analysis of the NVD, systematically evaluating each data field it contains.
%
In particular, we identify significant data issues with the vulnerability publication date, affected vendor and product names, severity scores, and vulnerability type.
We quantify the scope of each issue within the NVD, providing an understanding of each issue's ramifications. Then, we develop accurate and automated
methods of correcting the information, thus producing an improved
and more reliable NVD dataset for the security community to use. We will be open-sourcing the tools we created for correcting the NVD data quality concerns, as well as the rectified dataset itself.
Finally, we perform several analysis case studies using our improved NVD. Beyond providing more reliable analysis results for core questions on vulnerability discovery, disclosure, and remediation, our case studies demonstrate how analysis conclusions and practical implications can greatly differ due to data quality issues.
Ultimately, this work will not only directly impact real-world security through an improved dataset used in practice, but highlight common pitfalls that can affect other sources of vulnerability information, providing lessons for improving them as well as their effective uses.

\BfPara{Applications and Implications} We show the pitfalls of using NVD by highlighting NVD's various inconsistencies and propose methods to fix them. Overall, the study can be utilized by the NVD towards the following end goals: 
\begin{enumerate*}
    \item The estimated disclosure date identification can enrich the vulnerability report for the end-user's perusal.
    \item The vendor and product inconsistency finding tool can be leveraged during the vulnerability reporting process to suggest suitable vendor and product names to analysts. Moreover, the observations from our analyses and measurements can used as a best practice when adding new vendors and product names in NVD.
    \item The deep learning-based CVSS v3 prediction engine can be leveraged by NVD and security analysts alike for uniform severity metric generation across the vulnerabilities in the database.
\end{enumerate*}

\BfPara{Contributions} (1) Through an extensive data-driven approach backed by web scraping, manual investigation, and machine learning-based automation, we assess the quality of NVD, identifying concerns affecting each vulnerability data field. (2) We identify methods to automatically remedy the data quality issues in NVD, providing a more reliable source of vulnerability information. (3) As case studies, we conduct several large-scale analyses of vulnerabilities, providing the most accurate findings to several basic but core questions on vulnerability discovery, disclosure, and remediation. (4) We shared the results of this work with the US National Institute of Standards and Technology, which maintains the NVD. Following that, NVD's schemas have been updated to remove the free-form vendor and product names that we identify as oft problematic~\cite{changeLog}.

\BfPara{Organization} We provide a review of the literature in \autoref{sec:related}, followed by an overview of the dataset in \autoref{dataset}. In \autoref{sec:MiscAnalysis}, we present our main study, followed by case study analyses in \autoref{sec:vul_analysis}, and a discussion in \autoref{sec:discussion}. We conclude our work in \autoref{sec:conclusion}.


\section{Related work}\label{sec:related}

\BfPara{Reliability of NVD} Quality issues in vulnerability databases, e.g., NVD, have been previously noted and studied. Nguyen and Massaci~\cite{NguyenM13} pointed out that the affected product versions in the NVD are often incorrect, observing that roughly 25\% of Google Chrome CVEs had an incorrect Chrome version string. Christey and Martin~\cite{christey2013buying} similarly explored issues in the NVD data and suggested reporting biases as a root cause.  Attila \etal~\cite{erdHosicommon} showed that CVSS metrics are more suitable for enterprise software products than personal ones. Dong \etal~\cite{DongGCX19} analyzed the inconsistencies in public security vulnerability reports, including the NVD, and found overclaims and underclaims in the affected software product versions. 

While these studies call attention to certain inconsistencies, our study stands out by providing a comprehensive and systematic investigation of incompleteness and inconsistencies across the NVD data fields. In addition to identifying and quantifying the data quality issues therein, we also develop methods for correcting them.

\BfPara{Vulnerability Analysis}
Our work provides vulnerability analyses using more consistent vulnerability information, thus expanding on the literature on vulnerability dynamics.

Previously, Shahzad \etal~\cite{ShahzadSL12} analyzed the vulnerability life cycle, and pointed out that remotely exploitable vulnerabilities represent 80\% of all of them. 
Earlier, Clark \etal~\cite{ClarkFBS10} outlined a relation between a product's familiarity and its first vulnerability disclosure: a shorter time between product release and first vulnerability discovery is shown for familiar products. Ozment and Schechter~\cite{OzmentS06} observed that 62\% of vulnerabilities in the OpenBSD system were {\em foundational} and took 2.5 years for them to be reported. 

Stock \etal~\cite{StockPL0R18} and Li \etal~\cite{LiDCKBMSP16} studied the vulnerability notification channels and their significance. Zhao \etal~\cite{ZhaoGL2015} empirically studied data from two web vulnerability discovery ecosystems for trend analyses.
Trinh \etal~\cite{TrinhCJ14} studied vulnerabilities in web applications. Saha~\cite{Saha2008} extended an attack graph-based vulnerability analysis framework to include complex security policies for efficient vulnerability analysis. 
Zhang \etal~\cite{ZhangCO11} used data from NVD to predict the time to next vulnerability, and argued that NVD provides poor predictions while pointing out inconsistencies, e.g., missing version information, release time, and other obvious errors. 
Votipka \etal~\cite{VotipkaSRHM18} suggested integrating hackers and improved security training for testers in the vulnerability discovery. Xiao \etal~\cite{XiaoSLLLD18} detected vulnerability exploitation at a 90\% rate. Sabottke \etal~\cite{SabottkeSD2015} proposed a Twitter-based detector to identify vulnerabilities likely to be exploited. Homaei and Shahriari~\cite{HomaeiS2017} analyzed vulnerability reports between 2008 and 2014 and observed that security professionals can prevent 60\% of them by focusing on only seven vulnerability categories. William \etal~\cite{WilliamsDBNHA18} proposed a framework to discover evolutionary patterns in the vulnerabilities.

\section{Dataset}\label{dataset}
We study the National Vulnerability Database (NVD)~\cite{nvd_web}, the U.S. government's repository of public vulnerability information, actively maintained by the National Institute of Standards and Technology (NIST). While there are other databases, we focused on the NVD because it is widely used (in part because it is public and free), and arguably serves as the industry standard for tracking vulnerabilities. Nonetheless, our exploration of the NVD can provide insights into using other vulnerability databases. For the NVD, reported vulnerabilities are analyzed and added in a standardized format. Specifically, NVD entries contain the following. (1) A Common Vulnerability Exposure (CVE) ID number~\cite{cve_mitre} that uniquely identifies the vulnerability. (2) The vulnerability entry's publication date. (3) The vulnerability type/category, as classified by the Common Weakness Enumeration (CWE)~\cite{cwef}. (4) The severity, as rated by the Common Vulnerability Severity Score (CVSS)~\cite{cvss}. Note that there are two CVSS versions, the historical CVSS v2 (v2) and the modern CVSS v3 (v3)~\cite{cvss3}, both on a scale from 0 to 10. Table~\ref{Table:v2v3labels} shows the CVSS severity level thresholds. Note that the v3 introduces a critical level of severity. (5) A list of vendors and products affected, as classified under the Common Platform Enumeration (CPE)~\cite{cpef}.  (6) Free-form vulnerability descriptions. There can be multiple descriptions, although the typical one explains the security concern. Another common description is a comment by the CVE entry evaluator. (7) Optionally, reference URLs (e.g., security advisories) are sometimes listed, providing vulnerability details.



\BfPara{NVD Scale} 
We use a snapshot of NVD captured on May 21, 2018. This snapshot includes 107.2K CVEs added to NVD over two decades (1998--2018). These vulnerabilities are categorized into 453 CWE types, affecting 18.9K vendors and 46.6K products. We observe that 37.5K recent CVEs have the modern v3 severity label, in addition to v2 labels, while the remaining CVEs only have v2 labels.

\begin{table}[t]
\begin{center}
\caption{Score thresholds of v2 \& v3 CVSS severity levels.}
\label{Table:v2v3labels}
\begin{tabular}{lcll}
\Xhline{2\arrayrulewidth}
\textbf{Label} &Abbreviation& \textbf{v2} & \textbf{v3}\\
\Xhline{2\arrayrulewidth}
None& -- & -- & 0.0\\
Low & (L) & 0.0--3.9 & 0.1--3.9\\
Medium & (M) & 4.0--6.9 & 4.0--6.9\\
High & (H) & 7.0--10.0 & 7.0--8.9\\
Critical & (C) & -- & 9.0--10.0\\
\Xhline{2\arrayrulewidth}
\end{tabular}
\end{center}\vs{5}
\end{table}

\section{Inconsistencies and Improvements}\label{sec:MiscAnalysis}
The quality of data in a vulnerability database can heavily impact vulnerability tracking and trend analyses. 
Prior work by Mu \etal~\cite{MuCYHXMW18} already identified that crowd-sourcing vulnerability information has limitations.
In this section, we analyzed the NVD CVE entries for inconsistencies and explored methods for rectifying them. We focused on assessing the standardized non-free-form fields, specifically the vulnerability's publication date, CWE class, CVSS rating, and the affected CPE. The remaining NVD fields (the vulnerability description and reference URLs) are free-form without a standardized structure, making it challenging to conceptually define and identify inconsistencies, which we leave for future investigation. Note that we focused on data consistency issues, not data error problems. We assumed that the data in the NVD is correct but perhaps represented inconsistently, such that one could identify the correct information without resorting to investigation beyond what is provided through the NVD.



\subsection{Publication Dates}
\label{sec:publication_dates}

\BfPara{Incompleteness}
Vulnerability analysis often depends on tracking when vulnerabilities became public.
For example, security analysts must consider how long a vulnerability has been public when prioritizing patching, calculating windows of exposure, or investigating incidents (such as in log analysis).
NVD records have a publication date, but this date only indicates when the entry was added to the database.
We observed cases where the NVD publication date does not give a clear picture of vulnerability.
For example, CVE-2011-0700 is a WordPress XSS vulnerability with an NVD publication date of March 14, 2011. However, the CVE entry includes a reference URL for a public security advisory disclosing the vulnerability over a month earlier on February 7, 2011.

\begin{table*}[ht]
\begin{center}
 \begin{threeparttable}
 
\caption{Common inconsistency patterns in vendor naming.}\label{tab:inconsistencyPatternStats}
\begin{tabular}{l|clllll|lllll}
\Xhline{2\arrayrulewidth}
\multirow{2}{*}{\textbf{Category}} & \multirow{2}{*}{\textbf{Tokens}} & \multicolumn{5}{|c|}{\textbf{Length(Longest Substring Match)$\geq$ 3}} & \multicolumn{5}{c}{ \textbf{Length(Longest Substring Match)$<$3}} \\ \cline{3-12} 
 &  & \multicolumn{1}{|c}{\#MP $=$ 0} & \multicolumn{1}{c}{\#MP $=$ 1} & \multicolumn{1}{c}{\#MP $>$ 1} & \multicolumn{1}{c}{Pref} & \multicolumn{1}{c|}{PaV} & \multicolumn{1}{c}{\#MP $=$ 0} & \multicolumn{1}{c}{\#MP $=$ 1} & \multicolumn{1}{c}{\#MP $>$ 1} & \multicolumn{1}{c}{Pref} & \multicolumn{1}{c}{PaV} \\ 
 \Xhline{2\arrayrulewidth}
Possible & \multicolumn{1}{r|}{260 (524)} & \multicolumn{1}{r}{78 (155)} & \multicolumn{1}{r}{319 (608)} & \multicolumn{1}{r}{6 (11)} & \multicolumn{1}{r}{293 (566)} & \multicolumn{1}{r|}{5 (10)} & \multicolumn{1}{r}{223 (381)} & \multicolumn{1}{r}{658 (1151)} & \multicolumn{1}{r}{18 (33)} & \multicolumn{1}{r}{2 (4)} & \multicolumn{1}{r}{2 (4)} \\
Confirmed & \multicolumn{1}{r|}{260 (524)} & \multicolumn{1}{r}{52 (103)} & \multicolumn{1}{r}{295 (561)} & \multicolumn{1}{r}{4 (7)} & \multicolumn{1}{r}{266 (513)} & \multicolumn{1}{r|}{3 (6)} & \multicolumn{1}{r}{53 (76)} & \multicolumn{1}{r}{201 (341)} & \multicolumn{1}{r}{11 (20)} & \multicolumn{1}{r}{2 (4)} & \multicolumn{1}{r}{2 (4)} \\
\Xhline{2\arrayrulewidth}
\end{tabular}
\begin{tablenotes}
\small
\item[1] The numbers outside the parentheses are unique vendor pairs, while the numbers inside are the names associated with them. 
\item[2] Considered inconsistency patterns: (1) identical names except for special characters (labeled as Tokens); (2) vendor names associated with identical product names (labeled as \#MP=X, where X is the number of matching product names), (3) one vendor name is a product of the other vendor name in the pair (labeled as PaV), and (4) one name is a string prefix of the other name (labeled as Pref). \item[3] For cases (2)--(4), the longest common substring (LCS) between names is used as a signifier ($|$LCS$|\geq3$ v. $|$LCS$|<3$). 
\item[4] Pairs with (\#MP=0 $\wedge$ $|$LCS$|=0$ $\wedge$ not Pref) are not included in this table, as they do not meet our vendor matching heuristics. \end{tablenotes}
\end{threeparttable}
\end{center}\vspace{-5mm}
\end{table*}

\BfPara{Identification and Improvement} We attempted to identify disclosure dates by leveraging the reference URLs. Li and Paxson~\cite{LiP17} and Anwar \etal~\cite{AnwarKNM2018} previously suggested approximating the disclosure date by mining these references, as many are web pages about the vulnerability and its publication date. 

We first extracted the domains from the URL references, finding that the 591.4K URLs in our data corresponded to 5,997 domains. We focused on the top 50 domains, covering more than 85\% of all URLs (we observed diminishing returns from considering additional domains). These top domains fall into three high-level categories: (1) other vulnerability databases (e.g., {\em {SecurtiyFocus}}), (2) bug reports or email archives threads (e.g., {\em Bugzilla}),
 and (3) security advisories (e.g., {\em {cisco.com}}). Note that some domains are not in English (e.g., {\em jvn.jp} is in Japanese). 
Each of the webpages may have a different structure. Thus, we built a separate crawler for each domain to extract the relevant publication date for the vulnerability information (if any). We note that 14 domains are no longer responsive (e.g., {\em{osvdb.org}} shut down in 2016). For a given CVE, we approximated its public disclosure date as the minimum of the dates extracted from the reference URLs or the NVD publication date.

\begin{figure} [t]
\centering
\includegraphics[width=0.44\textwidth]{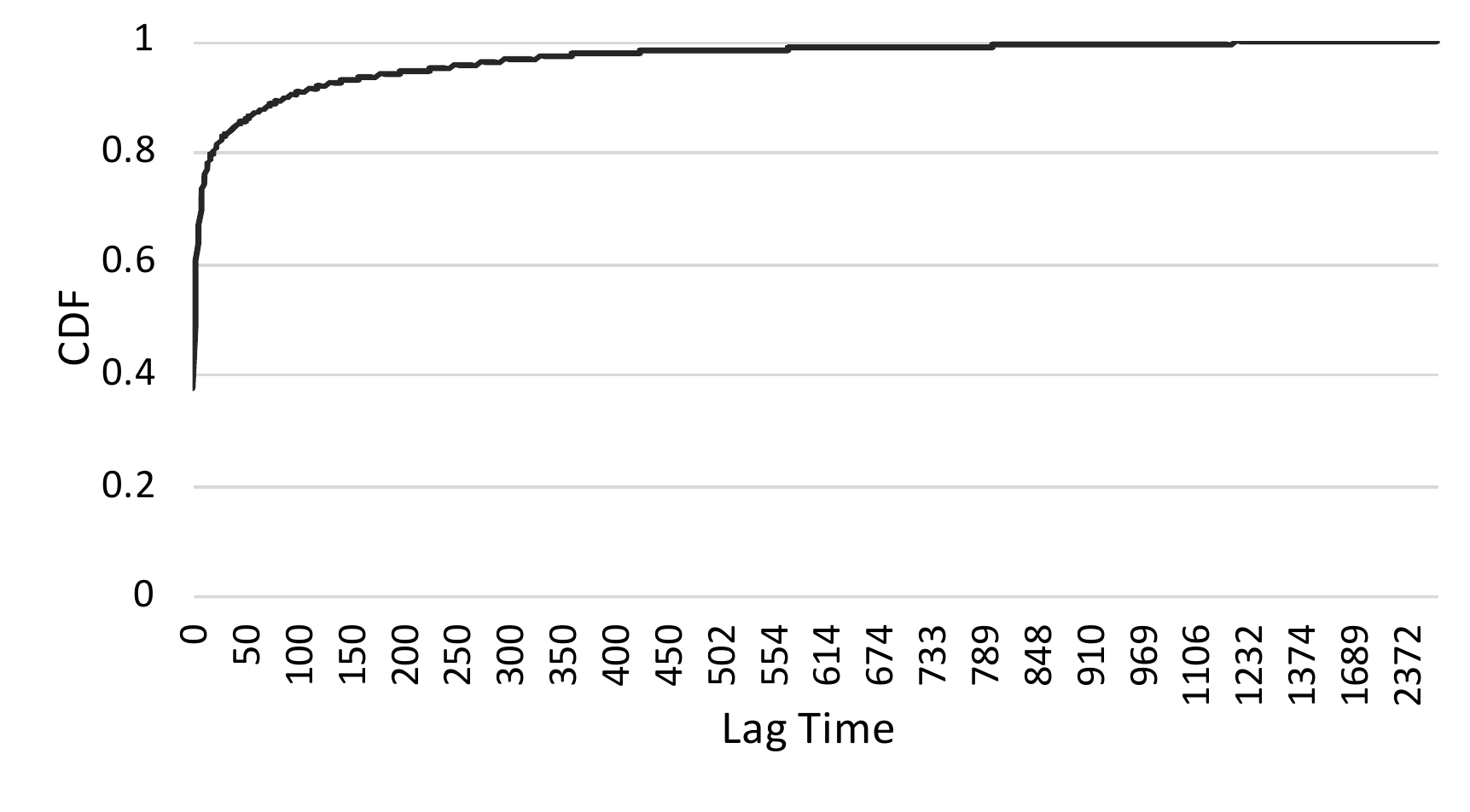}
\caption{CDF of vulnerability lag times. Lag time is the number of days after our estimated disclosure date when a vulnerability enters into the NVD. Note, $\approx$38\% of the vulnerabilities have no lag.}\vs{5}
\label{fig:lagTime}
\end{figure}

\BfPara{Improvement Impact} We evaluated how many days the CVE published date preceded our estimated disclosure date, which we call the lag time. \autoref{fig:lagTime} plots the percentage of CVEs within a lag time. Notice that $\approx$38\% of the vulnerabilities have a lag of zero days. The growth of vulnerabilities by lag time slows after accounting for the vulnerabilities with a lag of $\leq6$ days ($\approx$70\%). We observed that $\approx$ 28\% of the vulnerabilities have a lag of more than a week. Moreover, we distributed the lag among the v2 labels and observed that we improved on the publication date for only 37\% of low severity vulnerabilities, in comparison to 41\% medium and 65\% high severity vulnerabilities. This observation is particularly interesting as vulnerability tracking and analysis of high severity vulnerabilities are likely most valuable and can be most affected by this inconsistency.

\subsection{Vendor and Product Names}\label{sec:vendorIncons}

\BfPara{Inconsistencies} Practitioners depend on lists of vendors and products affected by a CVE to identify vulnerabilities affecting software they use~\cite{ShiraniCALD0H18}, or to monitor the security trends of various software systems. We observed inconsistencies in these vendor and product names.
For example, BEA Systems (vendor) is labeled as both {\em bea} (171 associated CVEs) and {\em bea\_systems} (14 different associated CVEs). 
Similarly, we observed AVG's anti-virus product has multiple names, including {\em antivirus} and {\em anti-virus}.
Thus, those monitoring for vulnerabilities by vendor or product names will obtain incorrect results unless carefully accounting for these inconsistencies.

\BfPara{Product Version Inconsistency} The NVD is also subject to inconsistent product versions, as demonstrated by Nguyen and Massaci~\cite{NguyenM13}.
Dong \etal~\cite{DongGCX19} leveraged NLP methods to find and correct inconsistencies in product versions through mining the NVD reference URLs. Thus, we did not investigate product versions further.

\BfPara{Identification and Improvement} Initially, we lack a general understanding of the nature of the vendor and product name inconsistencies. Thus, we resorted to manually analyzing name pairs to determine if both names represent the same entity (which we will call {\em matching pairs}). However, the manual analysis does not scale to the number of unique name pairs. To reduce the scale to a manageable level, we used heuristics to filter pairs down to those that are likely matching (i.e., related to the same entity yet with inconsistent names). We recognized that these heuristics should provide broad coverage but may not be truly comprehensive.

\BfPara{Vendor Names}
Informed by manual exploration, we developed three heuristics to identify likely matching vendor name pairs.
\begin{itemize}
\item Vendor name pairs share characters in common. This accounts for various scenarios such as where one name is misspelled (e.g., {\em microsoft} and {\em microsft}), represented in a different format (e.g., {\em avast} and {\em avast!}), abbreviated (e.g., {\em lan\_management\_system} and {\em lms}), or a strict substring of another (e.g., {\em lynx} and {\em lynx\_project}).

\item A product name is used as a vendor name (e.g., {\em microsoft} and {\em windows} both appearing as vendors).

\item Vendor pairs share the same product name.

\end{itemize}

\begin{table}[t]
\begin{center}
\caption{Vendor and product name inconsistencies in NVD, SecurityFocus (SF), and SecurityTracker (ST).}
\label{tab:inconsistencyStats}
\begin{threeparttable}
\begin{tabular}{l|rrr|rrr}
\Xhline{2\arrayrulewidth}
\multirow{2}{*}{\textbf{Database}} & \multicolumn{3}{c|}{\textbf{Vendor}} & \multicolumn{3}{c}{\textbf{Product}} \\
\cline{2-7}
 & \multicolumn{1}{c}{\#} & \#imp. & \#con. & \multicolumn{1}{c}{\#} & \#imp. &  \#ven. \\
\Xhline{2\arrayrulewidth}
NVD & 18,991 & 1,835 & 871 & 46,685 & 3,101 & 700 \\
SF & 24,760 & 2,094 & 878 & - & - & - \\
ST & 4,151 & 110 & 53 & - & - & - \\
\Xhline{2\arrayrulewidth}
\end{tabular}
\begin{tablenotes}
\small
\item $^1$ For both vendors and products, we list the number (\#) of distinct names and \# impacted by a discrepancy (\#imp). $^2$ For vendors, we list the number of consistent vendor names that map to inconsistent vendor names (\#con). $^3$ For products, we list the number of vendors affected by inconsistent product names. We only investigated produce names for the NVD. 
\end{tablenotes}
\end{threeparttable}
\end{center}\vspace{-7mm}
\end{table}

We filtered out vendor name pairs that do not satisfy any of these heuristics, and manually investigated each remaining pair by researching their products, developers, and associated organizations.
For each group of matching name pairs that represent the same vendor, we created a mapping of vendor names to consolidate those representing the same vendor under a consistent name. Note that there may be multiple matching pairs associated with the same vendor, indicating multiple inconsistent names. For the names associated with a vendor, we considered the one with the most associated CVEs as the consistent name, and remapped inconsistent vendor names in the NVD using our mapping.

To shed light on common patterns in inconsistent vendor naming, in Table~\ref{tab:inconsistencyPatternStats}, we listed those common patterns, as well as how likely those patterns signals a matching pair. We observed that 260 name pairs were identical except for the inclusion of special characters (e.g., ! or \_), and all were matching vendor name pairs. For other name pairs, when the longest substring match was at least 3 characters, the majority (at least 60\%) of name pairs were matching under the other patterns. Notably, when the two vendor names in the pair were both associated with the same product name, or when one vendor name was a string prefix of the other, the pair were matched in over 90\% of cases. When the longest substring match was less than 3 characters, only a minority of name pairs were still matching under the different patterns.

\BfPara{Product Names} 
After consolidating vendor names (above), we identified likely matching product names under the same (consolidated) vendor using two heuristics, and then manually evaluated the pairs. For the first heuristic, we tokenized product names by splitting by white spaces and special characters, and considered a product name pair as likely matching if the two tokenized names are identical. This captures cases such as {\em internet-explorer}, {\em internet_explorer}, and {\em internet explorer}. For the second heuristic, if one product name in the pair is tokenized into multiple components and the other is a single component, we concatenated the first character of the multi-component name, and compared the concatenated string with the other product name. This captures abbreviations, such as with {\em internet-explorer} and {\em ie}. Next, we investigated replacing, adding, and swapping of characters. We did so by determining the edit distance between product pairs. This is followed by manual verification of the pairs. The product names varying by characters can be different products altogether, e.g., {\em cisco}'s {\em ucs-e160dp-m1_firmware} and {\em ucs-e140dp-m1_firmware} have an edit distance of one, but are different products. With our analysis, we focused on pairs that can be a result of human error, e.g., {\em nativesolutions}'s {\em tbe_banner_engine} and {\em the_banner_engine}. As with vendor names, we mapped inconsistent product names to a consistent name based on the name associated with the most CVEs, and remapped product names in the NVD based on this mapping. Table~\ref{tab:inconsistencyStats} depicts that we found over 3K products inconsistently named affecting 700 vendors.

We note these two heuristics are more limited than those considered for vendor names, as we found that product names are often quite similar without representing the same product. For example, we explored using substring matching heuristics (as with vendor names), but found the number of pairs flagged for analysis to be too large and with many false positives (i.e., non-matching pairs).

\BfPara{Improvement Impact} Table~\ref{tab:inconsistencyStats} lists the extent of the vendor and product naming inconsistencies we identified. The NVD includes $\approx$19K distinct vendors, and about 10\% of them were impacted by vendor naming inconsistencies. These $\approx$1.8K vendor names could be consolidated under 871 vendor names, thus removing $\approx$5\% of distinct vendors.
Inconsistencies similarly affected 6\% of distinct NVD product names, and consolidating names would reduce the number of product names also by about 5\%. Thus, inconsistencies affect a non-trivial fraction of vendors and products. These numbers are lower bounds on the extent of vendor and product name inconsistencies in the NVD, since our identification and correction method relied on heuristics that may not be all-encompassing.

We also explored vendor naming inconsistencies in two other vulnerability databases with this information, SecurityTracker~\cite{sectrac}, and SecurityFocus~\cite{secfoc}. We used the same vendor name mapping that we generated (above) for correcting to consistent names, and applied it to the vendor strings in these two databases. As a result, we found as shown in Table~\ref{tab:inconsistencyStats} that 3\% and 8\% of vendor names were inconsistent for SecurityTracker and SecurityFocus, respectively. Exploration of these databases specifically will likely yield further inconsistencies, highlighting that this data quality issue is prominent in vulnerability database generally, and our approach for rectifying the NVD could be used for our datasets as well.

We note that Dong \etal~\cite{DongGCX19} also investigated product names specifically, where their heuristic was to split product names by white spaces into words, and label two products as matching if they shared words. In comparison, their method does not account for abbreviations or special character separators, and yield false positives when different products share similar words (e.g., Microsoft's {\em Internet Explorer} and {\em Internet Information Services} products).

\subsection{Severity Scores}\label{cvsssubsec}

\begin{table}[t]
\begin{center}
\caption{Transformation from v2 to v3 in numbers.}
\label{tab:v2Movements}
\scalebox{0.7}{
\begin{tabular}{l|rr|rr|rr|rr}
\Xhline{2\arrayrulewidth}
\multirow{2}{*}{\backslashbox{\textbf{v2}}{\textbf{v3}}}  & \multicolumn{2}{c|}{\textbf{L}} & \multicolumn{2}{c|}{\textbf{M}} & \multicolumn{2}{c|}{\textbf{H}} & \multicolumn{2}{c}{\textbf{C}} \\
\cline{2-9}
  & \multicolumn{1}{c}{\#}       & \multicolumn{1}{c|}{\%}         & \multicolumn{1}{c}{\#}        & \multicolumn{1}{c|}{\%}        & \multicolumn{1}{c}{\#}        & \multicolumn{1}{c|}{\%}        & \multicolumn{1}{c}{\#}        & \multicolumn{1}{c}{\%}        \\
  \Xhline{2\arrayrulewidth}
L & 363  & \acc{10}9.53  & 3,211 & \acc{84}84.30 & 235 & \acc{6}6.17 & 0 & 0.00      \\
M & 242 & \acc{1}1.07 & 10,589 & \acc{47}46.88 & 11,136 & \acc{49}49.30 & 621 & \acc{3}2.75 \\
H & 0 & 0.00 & 549 & \acc{5}4.96 & 5,293 & \acc{48}47.80 & 5,232 & \acc{47}47.24  \\
\Xhline{2\arrayrulewidth}
\end{tabular}}
\end{center}\vspace{-5mm}
\end{table}

\BfPara{Inconsistencies}
\label{sec:cvss_inconsistencies}
NVD uses the CVSS standard for rating severity~\cite{cvss}. However, CVSS has had multiple versions, with the modern v3 addressing limitations of prior versions. As v3 was only released in 2015, only a third of the CVEs in our NVD dataset have v3 scores. Security analysts monitoring vulnerabilities over time must either rely on v2 and its limitations (e.g., inaccurate security ratings), or evaluate a subset of the NVD data. Vulnerabilities pre-dating the release of v2 are still relevant, as age-old vulnerabilities are often still used in active attacks. For example, CVE-2011-0997 (a DHCP client vulnerability) was disclosed in 2011 yet could be used to target Avaya desk and IP conference phones in 2019~\cite{avaya_attack}. Similarly, CVE-2004-0113 is a medium severity vulnerability under v2 that was actively exploited in 2018 (over 14 years after disclosure) to exploit hosts and install crypto-mining malware~\cite{oldVulnExploited}.
Thus, we would ideally be able to backport v3 scores throughout the NVD, providing a more modern security rating for all vulnerabilities.

\BfPara{Identification and Improvement} Identifying CVEs with only v2 is straightforward, as NVD entries list the CVSS version associated with a score. The challenge is then improving the NVD by automatically assigning v3 scores to all CVEs. Both CVSS versions are calculated from a weighted aggregation of an input set of feature values, with v3 providing additional features and refined weighting. Thus, our approach is to develop a machine learning model that inputs v2 features, as well as other CVE entry information, and outputs meaningful v3 scores (despite lacking explicit features that normally are input into the v3 calculations). To evaluate the accuracy, we aimed not to necessarily produce identical severity scores as v3 would output, but predict the correct severity category (low, medium, high, critical) as the v3 score, which is commonly used for vulnerability prioritization~\cite{cvss}. We specifically applied a machine and deep learning approaches to model the potentially complex weighting and interactions between different features.

\BfPara{Features} While most parameters required for the severity scores remain the same as in v2, the parameters in v3 capture an annotated impact by the vulnerability. For example, ``access vector'' in v2 was transformed into ``attack vector'' in v3 with the specific effect of vulnerability into Physical (P), Network (N), Adjacent (A), and Local (L) impacts. Where v2 considered P attacks  as L, v3 divides the scores and introduces a new scope parameter, for vulnerabilities impacts beyond the exploitable system. The access complexity in v2 was divided into attack complexity and user interaction in v3, although the influence of the temporal metric is decreased in v3. To this end, we used the following v2 parameters as features to extrapolate v3 scores: access vector and complexity, authentication, integrity, availability, all privilege, user privilege, and other privilege flags. 

Holm and Afridi~\cite{HolmA2015} studied CVSS reliability by surveying 384 experts and 3,000 vulnerabilities, concluding the reliability depends on the vulnerability type. Thus, we add CWE-ID to our features.

\BfPara{Ground Truth Dataset} For ground truth, we need a mapping of v2 to v3 scores (or categories). As such, we used the recent CVEs in the NVD with both CVSS versions ($\approx$37K CVEs). We note that changes in the v3 score emphasize a better expressiveness for vulnerabilities' impact. The effect of these changes on the vulnerabilities is summarized in~\autoref{tab:v2Movements}, with no significant change in population, i.e., no vulnerability moves from Low in v2 to Critical in v3 and no vulnerability moves from High in v2 to Low in v3.

\BfPara{Model's Training}
Using the aforementioned features, we predicted the v3 base scores for vulnerabilities that do not have the v3 metrics. We began by splitting the ground truth data into 80\% training and 20\% testing datasets evenly distributed among classes. Additionally, we observe non-linear patterns among the v2 and v3 relation (see~\ref{app:featurePattern} for details).  We then applied a range of machine and deep learning prediction algorithms to predict the v3 scores: (1) Linear Regression (LR), (2) Support Vector Regression (SVR), (3) Convolutional Neural Networks (CNN), and (4) Deep Neural Networks (DNN). Linear regression finds the linear relationship between a target and one or more features. In addition, we used Support Vector Machine (SVM) as a regression method to predict v3 base score; we conducted the prediction using various combinations of parameters and report the best performing model (kernel type = rbf (radial basis function), kernel coefficient = 0.1, and penalty parameter = 2). We leveraged different deep learning techniques to extract deep feature representations for the vulnerabilities. We implemented a CNN model consisting of four consecutive convolutional layers. The first two layers consist of 64 filters and the remaining layers consist of 128 filters with a filter size of $3\times3$. The convolutional layers are followed by a flattening operation and a fully connected layer with 512 neurons. Next, a single neuron with a sigmoid activation function is used to output the prediction of the model. The sigmoid activation function is defined as 
$f(x) = \frac{1}{1+e^{-x}}$.
Similarly, we implemented a DNN model consisting of four fully connected layers with size of 128, 128, 256, and 256, respectively. The fully connected layers are followed by a single neuron with a sigmoid activation function to output the prediction of the model. We trained the deep learning models over 100 epochs using mean squared error loss function, $\frac{1}{N}\sum_{i=0}^{N}(y(x_i) - f(x_i))^2$, and Adam optimizer with a learning rate of 0.001. 
For evaluation, we defined the average error (AE) as ${[\sum_{i=0}^{N}Abs(y(x_i) - f(x_i))]}/{N}$, where $x_i$ is the $i^th$ sample of the testing dataset, $y(*)$ is the v3 severity score of the sample, $f(*)$ is the predicted value of v3 severity score of the sample, and $N$ is the size of the testing dataset. Similarly, we defined the average error rate (AER) as ${[\sum_{i=0}^{N}Abs(y(x_i) - f(x_i))/y(x_i)]}/{N}$.

\begin{table}[t]
\begin{center}
\caption{Prediction results: Average error (AE) and AE Rate (AER).}
\label{tab:predictionResults}
\scalebox{0.89}{\begin{tabular}{l|rrrr}
\Xhline{2\arrayrulewidth}
\multicolumn{1}{c|}{\textbf{Algorithm}} & \multicolumn{1}{c}{\textbf{LR}} & \multicolumn{1}{c}{\textbf{SVR}} & \multicolumn{1}{c}{\textbf{CNN}} & \multicolumn{1}{c}{\textbf{DNN}} \\
\Xhline{2\arrayrulewidth}
AER (\%) & \acc{12}12.16 & \acc{12}12.63 & \textbf{9.62} & \acc{11}11.61 \\
AE & 0.73 & 0.82 & \textbf{0.54} & 0.65 \\
\Xhline{2\arrayrulewidth}
\end{tabular}}
\end{center}\vspace{-5mm}
\end{table}

\BfPara{Model Learning Results} \autoref{tab:predictionResults} shows the average error and error deviation for different machine learning algorithms. The table shows that CNN has the lowest error rate and average error.
\autoref{tab:predictionAccuracy} shows the overall accuracy of our prediction engine. The overall accuracy of 86.29\% means that our model cannot predict the v3 scores correctly for 13.71\% of the vulnerabilities. These 13.71\% vulnerabilities were not correctly characterized by the v2 but are correctly characterized by v3. Moreover, we translated the v3 base scores to their respective severity labels according to the ranges in \autoref{Table:v2v3labels}. \autoref{tab:predictionAccuracy} lists the accuracy per input class, and we found that the model performs best for the input class High, i.e., with 93.55\% accuracy, and performs worst for target class Low, i.e., with 82.84\% accuracy. However, we also observed that DNN performs slightly better than CNN for the input class Low. Furthermore, we also tried other machine learning algorithms, and found that deep learning-based models (CNN and DNN) outperformed those alternatives. Given that the CNN-based model outperforms DNN-based model by $\approx$2\%, overall, we chose the CNN-based model for prediction.

\BfPara{Improvement Impact}  With our model, we can assign v3 scores and severity levels to all vulnerabilities in the NVD. For over 74K CVEs with only v2 scores, Table~\ref{tab:v2MovementsPV3} depicts their severity categories under v2 and our predicted v3. We observed that 48K CVEs change severity levels under v3, with 29K CVEs changing severity categories if we consider v2 High and v3 Critical to be equivalent (as v2 lacks a Critical level). Thus, nearly 40\% of CVEs have different severity once the severity score is updated to v3. Overall, the change skews towards high severity ratings. We hypothesized this is because v3 was designed in part to account for the scope of software affected, which can elevate the severity of a vulnerability when other sensitive systems are involved beyond the system immediately vulnerable. As a result, users of the NVD can prioritize better the vulnerabilities that they analyze and address.

The most impacted vulnerabilities by v3 do not adhere to any patterns, as confirmed from the prediction results, highlighting the power of our learning techniques in capturing complex mappings (see Appendix~\ref{app:featurePattern} for detailed analysis). 
Note that both old vulnerabilities mentioned earlier that are still exploited are more properly categorized as critical severity under our model (whereas one was medium severity and the other was high severity, with v2 labels).

\begin{table}[t]
\begin{center}
\caption{The v2 and v3, where v3 labels are  predicted by our model.}
\label{tab:v2MovementsPV3} 
\scalebox{0.7}{
\begin{tabular}{l|rr|rr|rr|rr}
\Xhline{2\arrayrulewidth}
\multirow{2}{*}{\backslashbox{\textbf{v2}}{\textbf{v3}}}  & \multicolumn{2}{c|}{\bf L} & \multicolumn{2}{c|}{\bf M} & \multicolumn{2}{c|}{\bf H} & \multicolumn{2}{c}{\bf C} \\
\cline{2-9}
  & \multicolumn{1}{c}{\#}       & \multicolumn{1}{c|}{\%}         & \multicolumn{1}{c}{\#}        & \multicolumn{1}{c|}{\%}        & \multicolumn{1}{c}{\#}        & \multicolumn{1}{c|}{\%}        & \multicolumn{1}{c}{\#}        & \multicolumn{1}{c}{\%}        \\
  \Xhline{2\arrayrulewidth}
L & 183  & 3.42  & 5,160 & 96.43 & 8 & 0.15 & 0 & 0.00      \\
M & 1 & 0.00 & 15,272 & 39.79 & 23,107 & 60.21 & 0 & 0.00 \\
H & 0 & 0.00 & 490 & 1.64 & 10,135 & 33.89 & 19,281 & 64.47  \\
\Xhline{2\arrayrulewidth}
\end{tabular}}
\end{center}\vspace{-5mm}
\end{table}

\begin{table}[t]
\begin{center}
\caption{Prediction accuracy. The overall accuracy of our prediction engine, and its accuracy by input class.}
\label{tab:predictionAccuracy}
\scalebox{0.9}{\begin{tabular}{l|cccc}
\Xhline{2\arrayrulewidth}
\multicolumn{1}{l|}{\multirow{2}{*}{\textbf{Accuracy}}} & \multicolumn{1}{c|}{\textbf{Overall}} & \multicolumn{3}{c}{\textbf{By input (v2) class (\%)}}                                                               \\ 
\cline{2-5}
\multicolumn{1}{l|}{}                  & \multicolumn{1}{c|}{(\%)}                         & \multicolumn{1}{c}{L}   & \multicolumn{1}{c}{M}    & \multicolumn{1}{c}{H}     \\ 
\Xhline{2\arrayrulewidth}
LR                                & \multicolumn{1}{r|}{\acc{83}83.14}                     & \multicolumn{1}{r}{\acc{83}82.58} & \multicolumn{1}{r}{\acc{79}79.31} & \multicolumn{1}{r}{\acc{91}91.14}  \\
SVR                                & \multicolumn{1}{r|}{\acc{66}66.46}                     & \multicolumn{1}{r}{\acc{83}82.97} & \multicolumn{1}{r}{\acc{71}71.15} & \multicolumn{1}{r}{\acc{51}51.21}  \\
CNN                                & \multicolumn{1}{r|}{\textbf{\acc{86}86.29}}                     & \multicolumn{1}{r}{\acc{83}82.84} & \multicolumn{1}{r}{\textbf{\acc{83}83.31}} & \multicolumn{1}{r}{\textbf{\acc{94}93.55}}  \\
DNN                                & \multicolumn{1}{r|}{\acc{84}84.41}                     & \multicolumn{1}{r}{\textbf{\acc{83}83.10}} & \multicolumn{1}{r}{\acc{81}80.67} & \multicolumn{1}{r}{\acc{93}92.48}  \\
\Xhline{2\arrayrulewidth}
\end{tabular}}
\end{center}\vspace{-5mm}
\end{table}

Johnson \etal~\cite{JohnsonLEF2016} assessed the credibility of CVSS scoring using a Bayesian method and found that, except for a few dimensions, CVSS is reliable. By analyzing five databases, they argued that NVD is the most reliable with respect to CVSS quality. In conducting our v3 extrapolation, we also argued that the predicted labels will help users prioritize vulnerabilities better. In particular, we found that the confidentiality, base score, and integrity are important features that impact the performance of our prediction model, i.e., the degree of information disclosure, the cumulative score of the vulnerability, and the degree of impact on the integrity of the victim. 

Allodi \etal~\cite{AllodiBFB2018} evaluated information affecting severity assessment. Our work extends their findings by showing which features determine the CVSS severity v3 score of a vulnerability. 



\subsection{Vulnerability Types}\label{vulntypesubsec}

\BfPara{Inconsistencies}  In the NVD, a CVE should be assigned a vulnerability type under the CWE classification~\cite{cwef} to provide users with an overview of the vulnerability nature and risk. Security analysts and developers leverage the vulnerability type to understand attack vectors that may impact their software, types of defenses to deploy, and track shifts in security concerns over time~\cite{cweFaq}. However, we identified that the CWE field for CVEs is not consistently populated correctly with a CWE-ID value. 

We found CVEs without CWE values, as well as 
those with their CWE entry as {\em NVD-CWE-Other}. By itself, this is missing data---rather than inconsistent, and out of the scope of our investigation (although worth noting for those analyzing NVD vulnerability types). However, we observed that the free-form CVE description (particularly the description provided by one of the vulnerability's evaluators) often contains the CWE-ID. For example, CVE-2007-0838 lists {\em NVD-CWE-Other} as its CWE-ID, while its evaluator description includes ``CWE-835: Loop with Unreachable Exit Condition ('Infinite Loop')''.  We also observed CVEs that list additionally relevant CWE-IDs in the description beyond those listed in the CWE field. In these cases, the CWE information is accessible in the CVE entry, but inconsistently provided.

\BfPara{Identification and Improvement} The CWE-ID follows a standard and distinct format that allows us to easily identify IDs in description strings through a regular expression (i.e., {\em CWE-[0-9]*}). For all CVEs, we applied this regular expression to the description strings to extract any CWE-IDs and add them to the set of CWE-IDs listed in the CWE field, if any. From this set of CWE-IDs, we filtered any CWE-ID values that indicate missing or non-specific CWEs (e.g., {\em NVD-CWE-Other}). In theory, descriptions could list CWE-IDs that are not relevant to the CVE (e.g., if discussing another vulnerability). However, through manually inspecting a random sample, we did not observe any erroneous cases where the CWE-ID in the description is not correct. Evidently, the CVE description outlines the traces of a vulnerability, which can be used to determine the type of vulnerability. We, therefore, investigated the capability of the CVE descriptions to extrapolate their corresponding types. We did so by utilizing different Natural Language Processing, machine learning, and deep learning techniques.

The crowd-sourced nature of the vulnerabilities devoid the descriptions of a standard descriptive pattern. Therefore, we began by preprocessing the data. Particularly, we unified the cases (convert text to lower case), removed the stop words and special characters (commonly used words that do not affect the meaning of the sentence, e.g., {\em This capability can be accessed} is changed to {\em capability access}), replaced contractions (e.g., {\em identifier's} is changed to {\em identifier}), and tense (past tense is changed to present tense, e.g., {\em used} is changed to {\em use}). Then, Universal Sentence Encoder~\cite{use_google}, a pre-trained transformer that is used to transform the text into high dimensional vector representation depending upon the semantic similarities and clustering, is utilized to represent the descriptions as vectors of size $1\times512$. The encoded vectors are then used to train and evaluate several machine learning and deep learning techniques, namely, k-Nearest Neighbor (k-NN), CNN, and DNN. We observed that k-NN (k = 1) provides the best results, predicting 151 different types with 65.60\% accuracy. While the results seem high considering the number of target classes, they cannot be reliably used given the criticality of the application.

\BfPara{Improvement Impact} By applying our CWE-ID extraction from CVE descriptions and matching CWE-ID name from the CWE list from their website~\cite{cweList}, we correct the CWE field for 2,456 vulnerabilities that do not have their types labeled. These vulnerabilities also include those that already have types assigned. Statistically, the existing database includes 26,312 vulnerabilities with NVD-CWE-Other label, 7,566 with NVD-CWE-noinfo label, and 1,293 with no assigned label, aggregating to $\approx$31\% of all the vulnerabilities. Additionally, we observed that most of the affected CVEs after our inconsistency fixes are those of type NVD-CWE-Others. Our analysis finds appropriate labels for 1,732 of the NVD-CWE-Other vulnerabilities and 14 of both the NVD-CWE-noinfo and unassigned vulnerabilities, making up for $\approx$5\% of those vulnerabilities.

\section{Case Studies}
\label{sec:vul_analysis}
With an improved and more consistent NVD, we conduct several vulnerability analyses as case studies on the impact of our NVD corrections.
For each analysis, we describe what questions are being asked, how the answers might be valuable in practice, the results from the analysis using both the original and rectified NVD data, and the impact of our improvements on the analysis outcome.

We recognize that there are a variety of potential analysis directions. This subset is by no means comprehensive, but rather involves informative questions one might reasonably ask when using the CVE fields we investigated from the NVD. While we believe the results of our analysis are useful for the security community, the ultimate goal of these case studies is to demonstrate how analysis results can be affected by the NVD data issues that we correct.





\begin{table}[t]
\begin{center}
\caption{Top 10 dates with the most vulnerabilities by CVE publication and our estimated disclosure dates (EDD). Day of week (DoW) and percent of that year's vulnerabilities reported on date are used.}
\label{NvdDateVuln}
\scalebox{0.84}{\begin{tabular}{c|c|c|c|c|c|c|c}
\Xhline{2\arrayrulewidth}
\multirow{2}{*}{\textbf{CVE Date}} & \multirow{2}{*}{\textbf{DoW}} & \multicolumn{2}{c|} {\textbf{Vulns}} & \multirow{2}{2em}{\textbf{EDD}} & \multirow{2}{*}{\textbf{DoW}} & \multicolumn{2}{c}{\textbf{Vulns}}\\
\cline{3-4}\cline{7-8}
                        & & \# & \% & & & \# & \%  \\
\Xhline{2\arrayrulewidth}
12/31/04	& F &	1,098	&	\acc{45}44.8	& 09/09/14 & T	&	384	&	\acc{5}5.1	\\
05/02/05	& M &	816	&	\acc{17}16.6	& 07/09/18 & M	&	359	&	\acc{3}2.4	\\
12/31/02	& T &	441	&	\acc{21}20.5	& 04/02/18 & M	&	344	&	\acc{2}2.3	\\
12/31/03	& W	&   407	&	\acc{27}26.7	& 07/05/17 & W	&	313	&	\acc{2}2.4	\\
07/09/18	& M	&   423	&	\acc{3}2.8	& 01/19/16 & T	&	295	&	\acc{5}4.6	\\
12/31/05	& Sa &	384	&	\acc{8}7.8	& 07/18/17 & T	&	275	&	\acc{2}2.2	\\
02/15/18	& Th &	340	&	\acc{2}2.3	& 07/14/15 & T	&	268	&	\acc{4}3.7	\\
09/09/14	& T &	326	&	\acc{4}4.1	& 05/02/05 & M	&	256	&	\acc{5}5.4	\\
08/08/17	& T &	316	&	\acc{2}2.2	& 01/17/17 & T	&	251	&	\acc{2}2.0	\\
04/18/18	& W &	281	&	\acc{2}1.9	& 07/17/18 & T	&	245	&	\acc{2}1.7	\\
\Xhline{2\arrayrulewidth}
\end{tabular}}
\end{center}\vspace{-3mm}
\end{table}

\begin{figure} [t]
\centering
\includegraphics[width=0.44\textwidth]{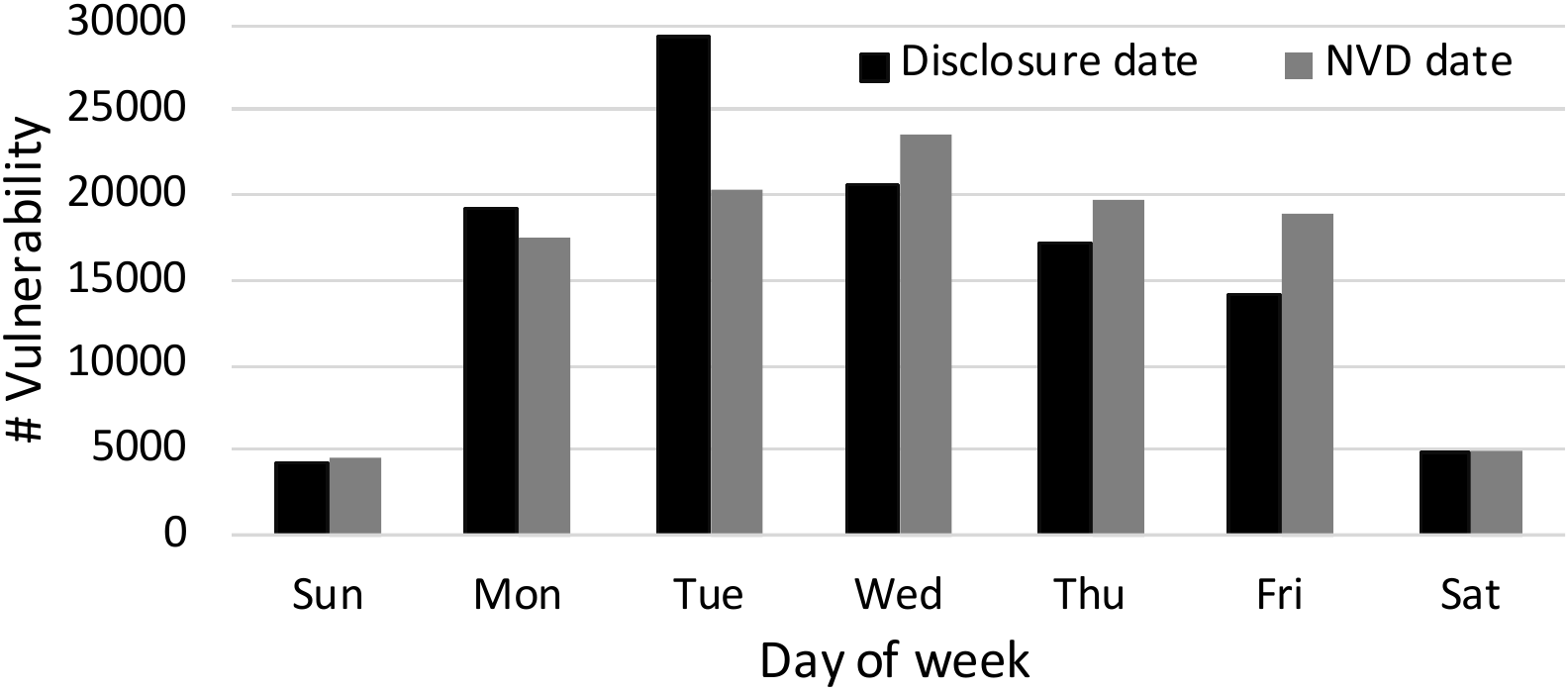}
\caption{The number of CVEs disclosed per day of the week (using our estimated disclosure dates) and published to the NVD.
}
\label{fig:dayofweek}\vspace{-5mm}
\end{figure}

\subsection{Vulnerability Disclosures}

\BfPara{Question} {\em When are vulnerabilities most frequently disclosed?}

\textbf{Analysis Value:} 
Understanding the times associated with high levels of vulnerability disclosures could shed light on underlying decisions in the disclosure process, as well as the impact of those decisions. For example, hypothetically, vendors could opt to disclose vulnerabilities at the end of the week or near holidays. As many people (including those working for media organizations) are off of work during subsequent periods, the vulnerabilities may draw less negative attention. As a consequence though, vulnerability remediation may be substantially delayed. It is important to understand if this indeed happens frequently.

\textbf{Analysis Results:}
\autoref{NvdDateVuln} shows the top 10 dates in terms of the number of vulnerability disclosures (based on our estimated disclosure date), as well as the day of the week for each date. When considering US holidays, we do not notice any particular pattern of pre-holiday disclosures. Rather, several of these top dates are within a couple of weeks after a US holiday, such as Independence Day (7/9/18, 7/5/17, 7/18/17, 7/14/15, and 7/17/18), Labor Day (9/9/14), and New Year's Day (1/17/17 and  1/19/16). Additionally, we note that these dates are primarily on Mondays and Tuesdays. To investigate this observation more broadly, \autoref{fig:dayofweek} shows the number of vulnerabilities disclosed on each day of the week. We find that beyond the top 10 dates, vulnerabilities are most frequently disclosed in the first half of a week (with fewer disclosures on Friday or over the weekend). In this analysis, we consider US holidays as most vendors in the NVD are US-based companies. However, we recognize that other nations celebrate many other holidays, and leave a more detailed global analysis for future work. We note that most vulnerabilities are disclosed during reasonable periods, where security professionals can obtain and act on information promptly.

\textbf{Impact of NVD Data Issues:}
For top CVE publication dates from \autoref{NvdDateVuln}, we observe New Year's Eve as four of the top 10 most active days, whereas it does not appear anywhere among the top 10 dates by our estimated disclosure dates. Most notably, on 12/31/2004, over 1K CVEs were added to the NVD, accounting for over 44\% of CVEs for that year. Yet according to our estimated disclosure date, only 175 were publicly disclosed that day. This discrepancy suggests an NVD artifact where a large number of CVEs may be added to the database before a new year arrives, or backdated to the last day of a prior year, rather than a more fundamental aspect of vulnerability reporting. Using the raw NVD data for vulnerability frequency analysis could produce inaccurate conclusions such as high vulnerability reporting during holidays. Similarly, \autoref{fig:dayofweek} indicates a more equal distribution of CVE publication dates throughout the week, which would incorrectly suggest many CVEs are indeed disclosed near weekends.



\begin{figure*}[t]
    \centering {\includegraphics[width=0.8\textwidth]{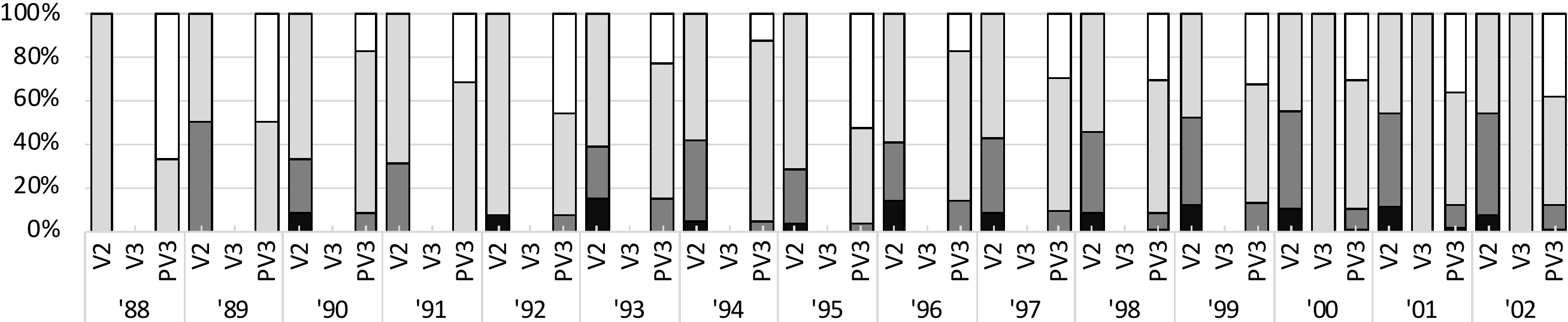}} {\includegraphics[width=0.8\textwidth]{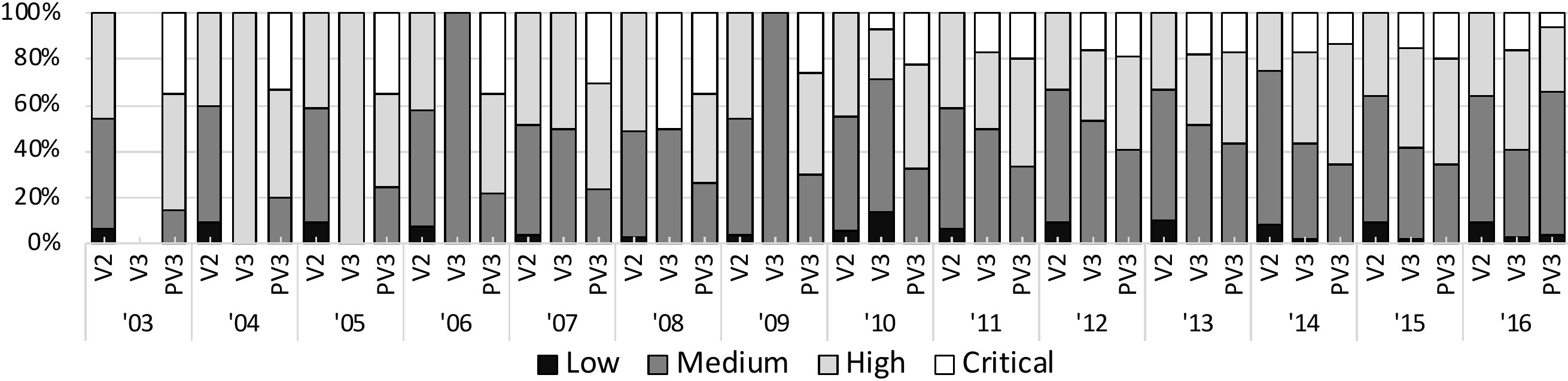}}\vs{2}
\caption{CVEs Distribution across severity categories over the years with different severity  scoring methods; v2, v3, and pv3 (our predicted v3 scores applied to all CVEs in the NVD; \textsection\ref{sec:cvss_inconsistencies}). Recall that v3 was only released in 2015, and all CVEs after 2017 were labeled with v3 scores. However, a subset of CVEs before 2017 was retroactively labeled with v3 scores.
}
\label{fig:lagOverSeverityClass}\vspace{-3mm}
\end{figure*}
\subsection{Vulnerability Severity}\label{severityImpact}

\BfPara{Question}  {\em What is the severity distribution of vulnerabilities?}

\textbf{Analysis Value:} 
As thousands of vulnerabilities are identified annually, it is vital that security practitioners can prioritize the most severe ones first. Furthermore, understanding what fraction of vulnerabilities receives each severity label allows them to identify how many vulnerabilities they may need to contend with. For the security community, it is also valuable to understand whether disclosed vulnerabilities skew towards low or high severity ones, shedding light on the nature of vulnerabilities being uncovered.

\textbf{Analysis Results:}
Recall that in Section~\ref{cvsssubsec}, we augmented the NVD by automatically applying accurate v3 severity ratings to all CVEs, rather than just relying on the most recent CVEs reported since v3 became standard.
In \autoref{version2_3Metrics}, we present the distribution of CVE severity (across all CVEs in the NVD) for both v2 and our predicted v3. In total, 8.25\% of all CVEs are low severity under v2, with the majority as medium severity. In contrast, under our predicted v3, less than 2\% are low severity, and the severity distribution is skewed towards the higher end, with the majority of vulnerabilities as high or critical severity. From both the v2 and v3 distributions, the small proportion of low severity vulnerabilities suggests some bias against discovering, reporting, or disclosing less urgent security concerns. However, v3's skew towards high severity ratings could spur different vulnerability remediation behavior, as many vulnerabilities rated as medium under v2 but higher under v3 might have been ignored by security practitioners earlier. 

\autoref{fig:lagOverSeverityClass} further breaks down the yearly distribution of CVEs across different severity categories, for v2, v3, and our predicted v3. Using our predicted v3 severity scores, we observe a decreasing trend in the proportion of critical severity CVEs over the years. For example, from 2011 onwards, less than 20\% of each year's CVEs were critical, compared to the early 2000s where nearly 30-40\% were likewise. This change indicates that the severity distribution of vulnerabilities is shifting over time. While we are uncertain of the cause of this shift, one hypothesis is that the increasing use of program analysis and fuzzing tools may be producing larger vulnerability populations than before, but the number of critical ones remains similar, thus resulting in a smaller proportion. Future work could investigate this phenomenon in more depth.

\begin{table}[t]
    \centering
        \caption{CVSS severity score distributions over all CVEs. }
    \begin{tabular}{lrr}
        \Xhline{2\arrayrulewidth}
          \textbf{Label}   & \textbf{v2 (\%)} & \textbf{Predicted v3 (\%)} \\
        \Xhline{2\arrayrulewidth}
        Low  & \acc{8}8.25 & \acc{2}1.62 \\ 
        Medium  & \acc{55}54.83 & \acc{38}38.30 \\  
        High & \acc{37}36.92 & \acc{44}44.48 \\ 
        Critical & N.A. & \acc{15}15.60 \\
        \Xhline{2\arrayrulewidth}
    \end{tabular}\vs{2}
    \label{version2_3Metrics}\vspace{-4mm}
\end{table}

\begin{table}[ht]

\begin{center}
\caption{Top 10 vulnerability types by the number of critical or high severity CVEs using v2, v3, and our predicted v3 (pv3) scores.
}\label{tab:cwesBySeverity}
\scalebox{0.75}{
\begin{tabular}{lr|lr|lr|lr|lr}
\Xhline{2\arrayrulewidth}
 \multicolumn{2}{c|}{\textbf{v2}} & \multicolumn{4}{c|}{\textbf{v3}} & \multicolumn{4}{c}{\textbf{pv3}} \\ \Xhline{2\arrayrulewidth}
 \multicolumn{2}{c|}{\textbf{High}} & \multicolumn{2}{c}{\textbf{Critical}} & \multicolumn{2}{c|}{\textbf{High}} & \multicolumn{2}{c}{\textbf{Critical}} & \multicolumn{2}{c}{\textbf{High}} \\
 \Xhline{2\arrayrulewidth}
 Type & \# & Type & \# & Type & \# & Type & \# & Type & \# \\ \Xhline{2\arrayrulewidth}
 BO\textcolor{blue}{$^{1}$} & 6935 & BO\textcolor{blue}{$^{1}$} & 1221 & BO\textcolor{blue}{$^{1}$} & 3025 & SQLI\textcolor{blue}{$^{2}$} & 3420 & BO\textcolor{blue}{$^{1}$} & 4078 \\
 SQLI\textcolor{blue}{$^{2}$} & 4115 & SQLI\textcolor{blue}{$^{2}$} & 673 & PM\textcolor{blue}{$^{3}$} & 1497 & BO\textcolor{blue}{$^{1}$} & 1783 & PM\textcolor{blue}{$^{3}$} & 2096 \\
 PM\textcolor{blue}{$^{3}$} & 2581 & IV\textcolor{blue}{$^{4}$} & 323 & IV\textcolor{blue}{$^{4}$} & 1291 & CI\textcolor{blue}{$^{5}$} & 766 & CR$^{18}$ & 1802 \\
 IV\textcolor{blue}{$^{4}$} & 2070 & UaF\textcolor{blue}{$^{7}$} & 271 & AC\textcolor{blue}{$^{11}$} & 955 & PM\textcolor{blue}{$^{3}$} & 601 & IV\textcolor{blue}{$^{4}$} & 1749 \\
 CI\textcolor{blue}{$^{5}$} & 1463 & AC\textcolor{blue}{$^{11}$} & 247 & IE\textcolor{blue}{$^{14}$} & 683 & IV\textcolor{blue}{$^{4}$} & 447 & RM\textcolor{blue}{$^{6}$} & 1426 \\
 RM\textcolor{blue}{$^{6}$} & 1416 & PM\textcolor{blue}{$^{3}$} & 232 & IO\textcolor{blue}{$^{15}$} & 680 & PT\textcolor{blue}{$^{9}$} & 364 & IE\textcolor{blue}{$^{14}$} & 1180 \\
 UaF\textcolor{blue}{$^{7}$} & 712 & IA\textcolor{blue}{$^{10}$} & 190 & CSRF\textcolor{blue}{$^{16}$} & 671 & AC\textcolor{blue}{$^{11}$} & 362 & PT\textcolor{blue}{$^{9}$} & 1173 \\
 NE\textcolor{blue}{$^{8}$} & 702 & CD\textcolor{blue}{$^{12}$} & 125 & UaF\textcolor{blue}{$^{7}$} & 443 & RM\textcolor{blue}{$^{6}$} & 341 & CI\textcolor{blue}{$^{5}$} & 1168 \\
 PT\textcolor{blue}{$^{9}$} & 672 & CMD\textcolor{blue}{$^{13}$} & 114 & BoR\textcolor{blue}{$^{17}$} & 414 & NE\textcolor{blue}{$^{8}$} & 295 & CSRF\textcolor{blue}{$^{16}$} & 984 \\
 IA\textcolor{blue}{$^{10}$} & 666 & CI\textcolor{blue}{$^{5}$} & 108 & PT\textcolor{blue}{$^{9}$} & 360 & UaF\textcolor{blue}{$^{7}$}& 224 & NE\textcolor{blue}{$^{8}$} & 777 \\ \Xhline{2\arrayrulewidth}
\end{tabular}}
\flushleft \scriptsize  
\textcolor{blue}{$^{1}$}Buffer Overflow, 
\textcolor{blue}{$^{2}$}SQL Injection, 
\textcolor{blue}{$^{3}$}Permission Management, 
\textcolor{blue}{$^{4}$}Input Validation, 
\textcolor{blue}{$^{5}$}Code Injection, 
\textcolor{blue}{$^{6}$}Resource Management, 
\textcolor{blue}{$^{7}$}Use-after-Free,
\textcolor{blue}{$^{8}$}Numerical Error, 
\textcolor{blue}{$^{9}$}Path Traversal, \\
\textcolor{blue}{$^{10}$}Improper Authorization, 
\textcolor{blue}{$^{11}$}Access Control, 
\textcolor{blue}{$^{12}$}Credentials, 
\textcolor{blue}{$^{13}$}Command, \\
\textcolor{blue}{$^{14}$}Information Exposure, 
\textcolor{blue}{$^{15}$}Integer Overflow, 
\textcolor{blue}{$^{16}$}Cross-Site Request Forgery, 
\textcolor{blue}{$^{17}$}Buffer Over Read.
\end{center}\vspace{-5mm}
\end{table}

\textbf{Impact of NVD Data Issues:}
In NVD, all CVEs since 2017 are assigned v3 scores. However, no CVE before 1999 has an assigned v3 score, and before 2013, no more than 35 CVEs each year have a v3 score retroactively labeled (as v3 was officially released at the end of 2015~\cite{nvdNews}). This minority of CVEs with assigned v3 scores is too limited for many analyses. For example, as seen in \autoref{fig:lagOverSeverityClass}, CVEs with assigned v3 scores in certain years are unrepresentative of the likely real severity distribution. In 2000-2002, 2004-2006, and 2009, only one severity level appears for all CVEs with assigned v3 scores.
While security analysts could rely on v2 instead, v3 was explicitly designed to overcome limitations of v2. Thus, our predicted v3 affords comprehensive severity analysis across the entire NVD dataset. This historical perspective is particularly important as vulnerabilities remain viable for years after disclosure~\cite{oldVulnExploited}.

\subsection{Vulnerability Types}

\BfPara{Question} {\em Which vulnerability type has most critical vulnerabilities?}

\textbf{Analysis Value:} Understanding which vulnerabilities are associated with the most critical CVEs is useful for both security practitioners and researchers, allowing them to prioritize which tools or defense systems to invest in or investigate.

\textbf{Analysis Results:}
Our analysis involves the CWE and CVSS severity fields. In table ~\ref{tab:cwesBySeverity} we list the top 10 CWE categories by the number of high/critical severity CWEs, using v2, v3, and pv3 severity scores.
By both correcting CWE labels and using our predicted v3 scores, we identify that SQL injection has the most critical CVEs, with almost twice as many as the next vulnerability type (buffer overflows). Meanwhile, for high-but-not-critical CVEs, buffer overflows are most common, and SQL injection does not appear within the top 10. This suggests that when SQL injection vulnerabilities are identified, they are typically of the utmost severity.

\textbf{Impact of NVD Data Issues:}
Buffer overflow and SQL injection are consistently the most frequent types under v2, v3, and our PV3. However, we note that overall, the top 10 CWE types for our PV3 more closely resembles that of v2, compared to v3. For example, access control, command injection, and hard-coded credentials are in the top 10 v3 critical CVEs, but not in v2 or our PV3. Thus, our corrected NVD results appear more consistent than using the original CWE and v3 NVD labels.

\subsection{Vendor and Product Names}

\BfPara{Question} {\em Which vendors have most CVEs or vulnerable products?}
 
 \textbf{Analysis Value:} Analysts may inform their operation using the vulnerability impact information across vendors, e.g., which vendors to track for new vulnerabilities, or which products to analyze.
 
 \textbf{Analysis Results:}
\autoref{vuln_product} shows the top 10 vendors per the associated CVEs and affected products, as a count and a fraction of all CVEs and affected products associated with each vendor. The statistics are presented for before and after our NVD corrections, but we will use the post-correction values for our analysis. 

We observe that the top vendors represent a significant fraction of all CVEs and products. The top 10 vendors account for about 36\% of all CVEs
and 22\% of all products. Thus, the impact of CVE vulnerabilities is concentrated on a small set of vendors, with a long-tail of the remaining less-impact ones. It is also interesting to note that the top vendors by CVE count are quite different than those by the product count, with only 4 common vendors. This difference suggests that the concentration of CVEs among top vendors is not simply due to these vendors supporting a wide number of products. 

\textbf{Impact of NVD Data Issues:}
The impact of product and vendor name inconsistencies is less dramatic for this analysis, as ultimately the order of top vendors remains the same before and after corrections. However, the changes in vulnerability counts can be notable. For example, Oracle had over 100 more associated CVEs after our naming fixes, and Debian had 95 more CVEs.

\begin{table}[t]
\centering
\caption{Top 10 vendors per the number of associated CVEs and affected products, after and before name corrections (\# is a count and \% as a percent of CVEs or products associated with that vendor).
}\vs{2}
\label{vuln_product}
\begin{minipage}[t]{0.25\textwidth}
\scalebox{0.75}{
\begin{tabular}{l|rrrr}
\Xhline{2\arrayrulewidth}
\multirow{3}{*}{\textbf{Vendor}} & \multicolumn{4}{c}{\textbf{\# of CVEs}} \\
\cline{2-5}
                        & \multicolumn{2}{c}{\textbf{After}} & \multicolumn{2}{c}{\textbf{Before}} \\
                        \cline{2-5}
                        & \#          & \%          & \#           & \%   \\
\Xhline{2\arrayrulewidth}
Microsoft               & 6,602        & 6.16        & 6,597         & 6.15        \\
Oracle                  & 5,650        & 5.27        & 5,526         & 5.15        \\
Apple                   & 4,574        & 4.26        & 4,574         & 4.26        \\
IBM                     & 4,160        & 3.88        & 4,160         & 3.88        \\
Google                  & 3,934        & 3.67        & 3,933         & 3.67        \\
Cisco                   & 3,674        & 3.43        & 3,674         & 3.43        \\
Adobe                   & 2,869        & 2.68        & 2,869         & 2.68        \\
Linux                   & 2,275        & 2.12        & 2,254         & 2.10        \\
Debian                  & 2,275        & 2.12        & 2,180         & 2.03        \\
Redhat                  & 2,161        & 2.01        & 2,144         & 2.00        \\

\Xhline{2\arrayrulewidth}
\end{tabular}}
\end{minipage}~
\begin{minipage}[t]{0.25\textwidth}
\scalebox{0.75}{
\begin{tabular}{l|rrrr}
\Xhline{2\arrayrulewidth}
\multirow{3}{*}{\textbf{Vendor}} & \multicolumn{4}{c}{\textbf{\# of Products}}  \\
\cline{2-5}
& \multicolumn{2}{c}{\textbf{After}} & \multicolumn{2}{c}{\textbf{Before}} \\
\cline{2-5}
& \#          & \%          & \#           & \%          \\
\Xhline{2\arrayrulewidth}
HP                      & 3,067        & 6.73        & 3,083         & 6.60        \\
Cisco                   & 1,821        & 4.00         & 1,839         & 3.94        \\
IBM                     & 926         & 2.03        & 926          & 1.98        \\
Axis                    & 808         & 1.77        & 808          & 1.73        \\
Intel                   & 721         & 1.58        & 723          & 1.55        \\
Huawei                  & 701         & 1.54         & 707          & 1.51        \\
Lenovo                  & 579         & 1.27        & 579          & 1.24        \\
Oracle                  & 553         & 1.21        & 546          & 1.17        \\
Siemens                 & 510         & 1.12        & 534          & 1.14        \\
Microsoft               & 489         & 1.07        & 486          & 1.04   \\
\Xhline{2\arrayrulewidth}
\end{tabular}}
\end{minipage}\vspace{-4mm}
\end{table}

\begin{table}[t]
    \centering
        \caption{CVEs with mislabeled vendors/products by severity levels using v2 labels and our predicted v3 (pv3) labels.}\vs{2}
    \scalebox{0.9}{
    \begin{tabular}{l|rrrr}
        \Xhline{2\arrayrulewidth}
 & \multicolumn{2}{c}{\textbf{Mislabeled Vendor}} & \multicolumn{2}{c}{\textbf{Mislabeled Product}} \\
 \cline{2-5}
 & v2 & pv3 & v2 & pv3 \\
\Xhline{2\arrayrulewidth}
Low & 275 & 10 & 27 & 4 \\
Medium & 2,033 & 1,101 & 196 & 105 \\
High & 1,206 & 1,484 & 159 & 205 \\
Critical & NA & 919 & NA & 68\\ 
       \Xhline{2\arrayrulewidth}
    \end{tabular}
    }
    \label{mislabelledCve}\vspace{-3mm}
\end{table}

Even when the number of CVEs with a mislabeled vendor or product is small, the security risk can be high. In \autoref{mislabelledCve}, we consider all CVEs with the corrected vendor or product label, and break down their severity levels using v2 and our predicted v3. While only several thousand CVEs were mislabeled and subsequently corrected, over a third are high severity under v2 and a quarter are critical under our predicted v3. In total, nearly 1000 mislabeled CVEs are critically severe. A security analyst tracking a particular product or vendor could easily miss relevant severe vulnerabilities, putting their systems at risk. (After all, it only takes one missed vulnerability to permit a security situation, such as with Equifax~\cite{equifaxStruts}.)

\section{Discussion}\label{sec:discussion}

{\em The Need for a Reliable Vulnerability Database.} Given the wide range of applications of vulnerability databases, in both the industry and the research community, the reliability of the information present in them is of the utmost importance. However, some of the key takeaways of this work show that the information in NVD is inconsistent, as demonstrated by the associated quantification, thereby raising questions on NVD's reliability. The inconsistencies are shown to vary, including the delay between a vulnerability's disclosure and its publish date in the NVD, to its vendor and product name, to its severity metrics, to the vulnerability type.  With this work, by identifying the inconsistencies, we highlight the pitfalls of using NVD.  Given the non-uniform state of the vulnerable systems, inconsistencies in them require manual effort. We conducted a manual investigation and then utilized the efforts to build an automated system to identify inconsistencies. For others, we built automated tools that can be used to recover consistency. 

While the estimated disclosure date in this study fundamentally questions the completeness of the NVD, other fixes address NVD's inconsistency. It is argued that the reports listed in the reference links in NVD might not be public or known at the time of their insertion into the NVD. In addition, the vulnerability information can be modified multiple times, as it is the practice with incremental vulnerability reporting. The proposed approach can therefore be utilized to change the estimated disclosure date of the vulnerability during a modification, given such practices and operational caveats.

\BfPara{Root Cause of Inconsistencies} Understanding the root causes of the inconsistencies in NVD can help eliminating them. Our analyses provide various plausible explanations for the root causes of inconsistencies. For vendor/product inconsistencies, we noticed that they were clearly due to the incorrect naming conventions, using developers as vendors, due to vendor acquisitions, and typos by analysts. Among those root causes, the acquisitions are a dynamic root cause, and therefore are difficult to mitigate, while other causes can be addressed by standardizing a nomenclature. 

\begin{figure} [t]
\centering
\includegraphics[width=0.44\textwidth]{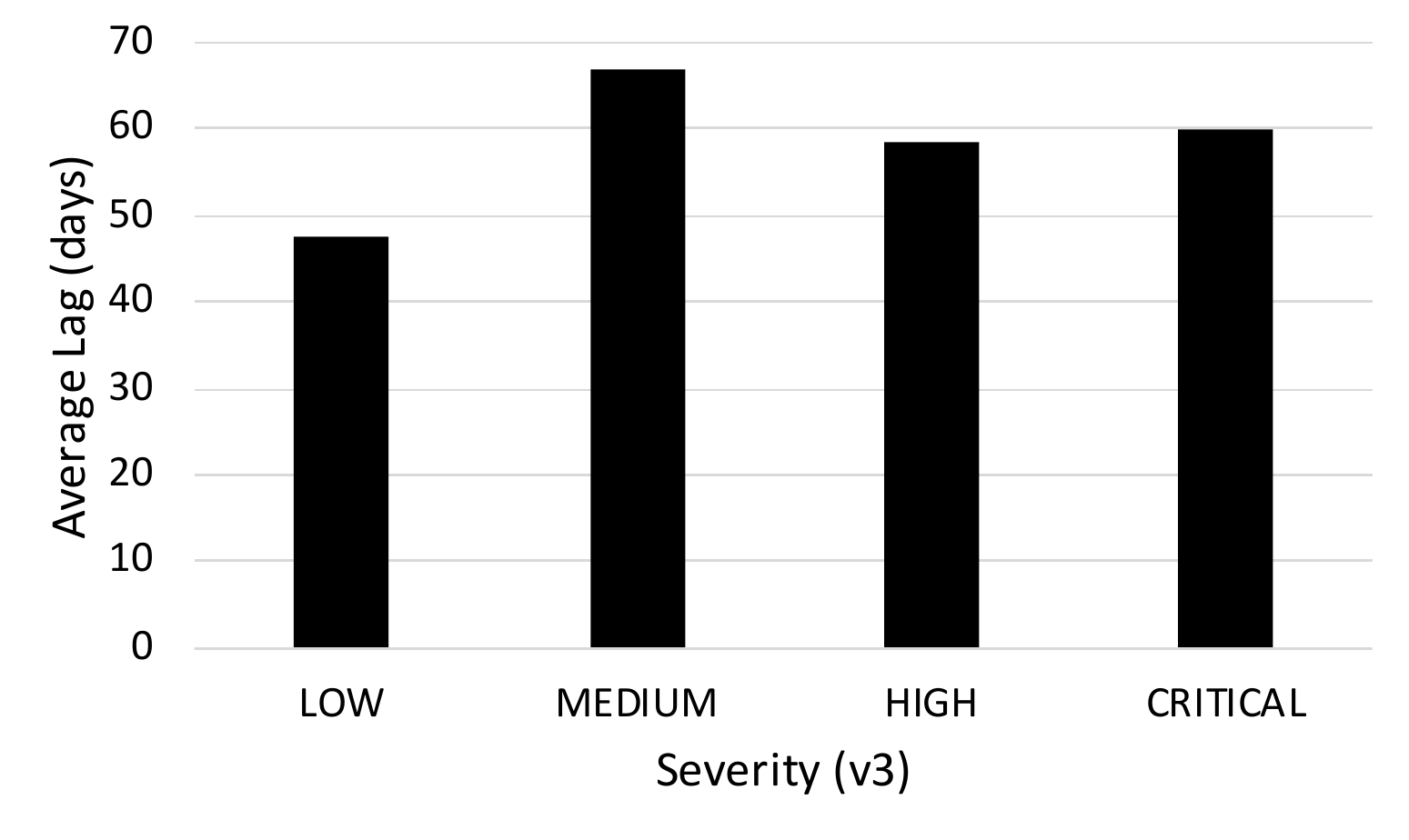}\vs{3}
\caption{Average lag time by v3 severity level.}
\label{fig:SeverityLag}\vspace{-5mm}
\end{figure}

The reason behind the inconsistencies in the v3 severity is the adoption of a new severity scoring system, which was not in existence at the time of scoring the severity of older vulnerabilities. Given the absence of the parameters that differentiate between v3 and v2, v3 was not generalized for those vulnerabilities, although such generalization was done by NVD when adopting v2 throughout with a considerable accuracy. Similarly, by leveraging the deep learning-based algorithms, we determined the v3 labels from the v2 labels. We investigated the severity of the vulnerabilities with a lag between the estimated disclosure date and the NVD date. \autoref{fig:SeverityLag} shows the average lag, in days, by the different severity levels in the v3, and we observe that the average among the various severity levels ranges between 47.6 days to 66.8 days, thereby demonstrating that the delay in the insertion of vulnerability into the NVD has no relationship with the severity of the vulnerability.

\BfPara{Applications} This work highlights inconsistencies in the NVD data fields, and proposes methods to fix them. The diversified inconsistencies warrant multiple tools, dealing with one at a time. As a result, this study can be utilized by the analysts at NVD towards the following goals: 
\begin{enumerate*}
    \item The estimated disclosure date identification can enrich the vulnerability report for the end-user's perusal. The tool enables the analysts to scrape through the different vulnerability reports and disclosures from the reference links of the recently added vulnerabilities and notify them of the disclosure date.
    \item The vendor and product inconsistency finding tool can be leveraged during the vulnerability reporting. The individual reporters can enter the vendor and product name according to their perception, and the tool will suggest the suitable vendor and product name from the generated consistent database. The reporter will then choose the consistent vendor and product name if available. Additionally, the NVD analysts can use the tool to re-assess the vendor and product names towards the generation of CPE URI (both 2.2 and 2.3). Moreover, for new vendor and/or product names, our observed inconsistencies and the root causes can help control the inconsistencies in the future (see Appendix~\ref{app:vendorConsisCaseStudy} for details).
    \item Our tool to determine the CVSS v3 metrics can be leveraged for a uniform severity metric across  vulnerabilities in the database. Moreover, it can be used by the users of NVD to prioritize their patching.
\end{enumerate*}

Leveraging the improved NVD, we formulate analysis questions as case studies to understand the impact of our corrective measures. Although there were numerous analyses that we came up with, we present the questions that a user might have when using the corrected fields. We observe that while public disclosures happen in the early days of the week, the inclusion of them in the NVD happens on the latter days. Additionally, the high reportage of CVEs on the last day of a year can be due to their retroactive inclusion when only the year was known. The temporal analysis of software weakness can help understand the trends to understand the up and the coming vulnerabilities. These emerging software weaknesses may be a result of a recently found attack vector. These can be utilized during the software product development and can help prioritize patching processes, and to emphasize upon, during the various phases of the software development life cycle. A consistent database would give a better picture of the trends, including their exploitation window (depending upon the disclosure date of a vulnerability and the date it is discovered on a host computer).

 

\BfPara{Limitations} To estimate the disclosure date, we consider the domain names representing 85\% of the URLs. The reduction of coverage by 15\% may lead to an imprecise estimation of the disclosure date. Moreover, vendor and product inconsistency numbers present a lower bound on inconsistencies that NVD may have. We would not group the vendors if another vendor acquired a probable inconsistent vendor. An approach to improve the bounds would require determining the date of acquisition of the probable inconsistent vendor and then correlating it with their estimated disclosure date. 

\section{Conclusion}\label{sec:conclusion}
Given the importance of such a database as NVD for security operations, identifying, measuring, and fixing the inconsistencies is essential, which we pursue through various tools, including multi-sourced web scraping, manual vetting, and deep learning algorithms for the publication date, vendor names, product names, severity categories, and vulnerability types inconsistency remedies. The inconsistency fixed database revealed exciting insights about the NVD and vulnerability reporting in general, and how basing the analysis on the current NVD leads to different conclusions than on the fixed one. The frequent days in estimated public disclosure and published date shows the prevalence of early days in the week (Monday and Tuesday) among disclosure dates and the latter days among publication date in the NVD. The fixed vendor names show decreasing inconsistencies over time, while product names need more attention for better resolution. The v3 fix reveals a better distribution of the v3 metric and the vulnerability type fix identifies additional types, other than the ones listed in the NVD.


{\normalsize \bibliographystyle{acm}
\bibliography{refBp,conf}}

\appendix
\section{Appendix}\label{app:1}
\subsection{Feature Pattern}\label{app:featurePattern}

\begin{figure*}[h]
    \centering
		\subfigure[Low\label{fig:cvssMovements:Low}] {\includegraphics[width=0.32\textwidth]{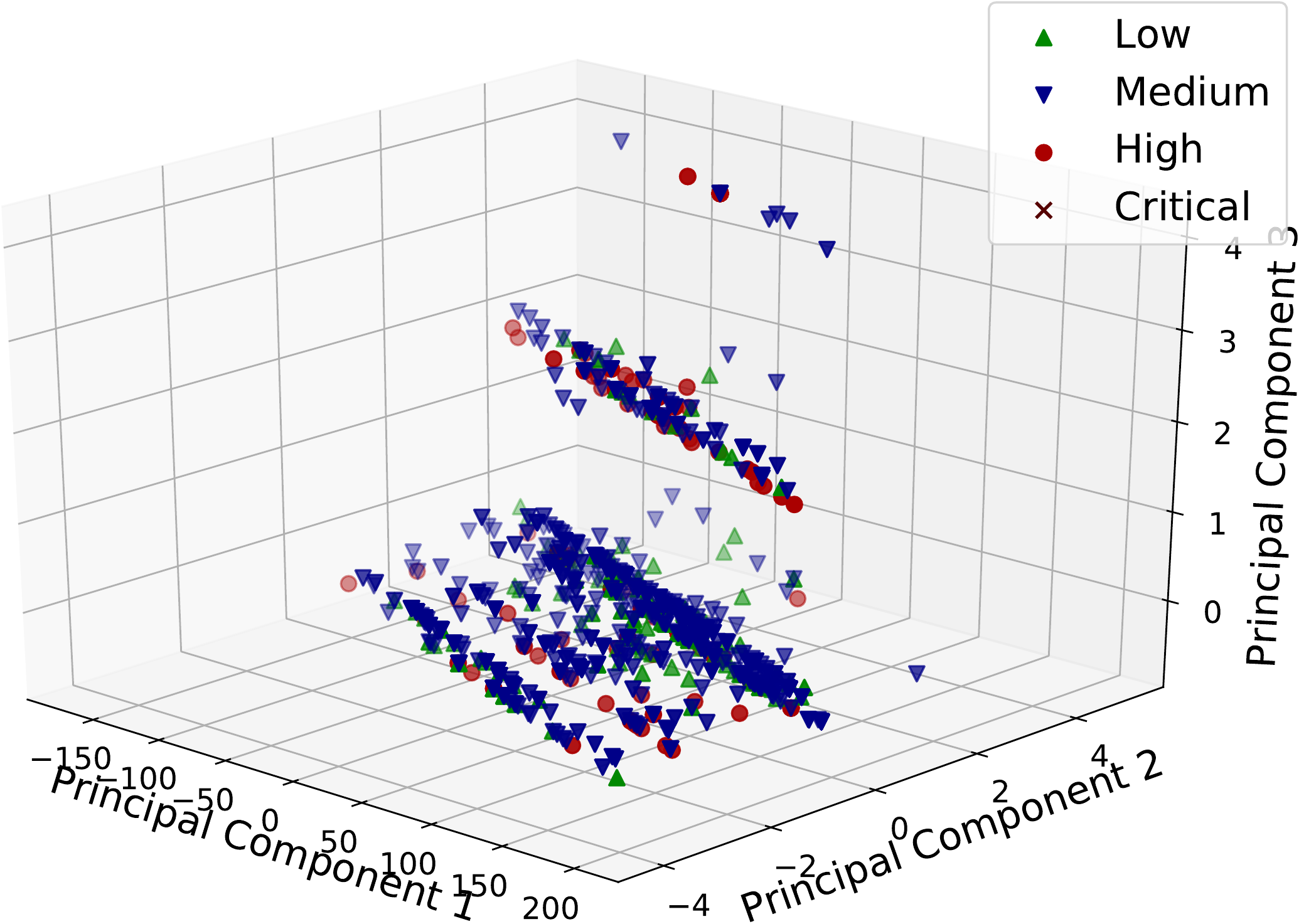}}
		\subfigure[Medium \label{fig:cvssMovements:Medium}] {\includegraphics[width=0.32\textwidth]{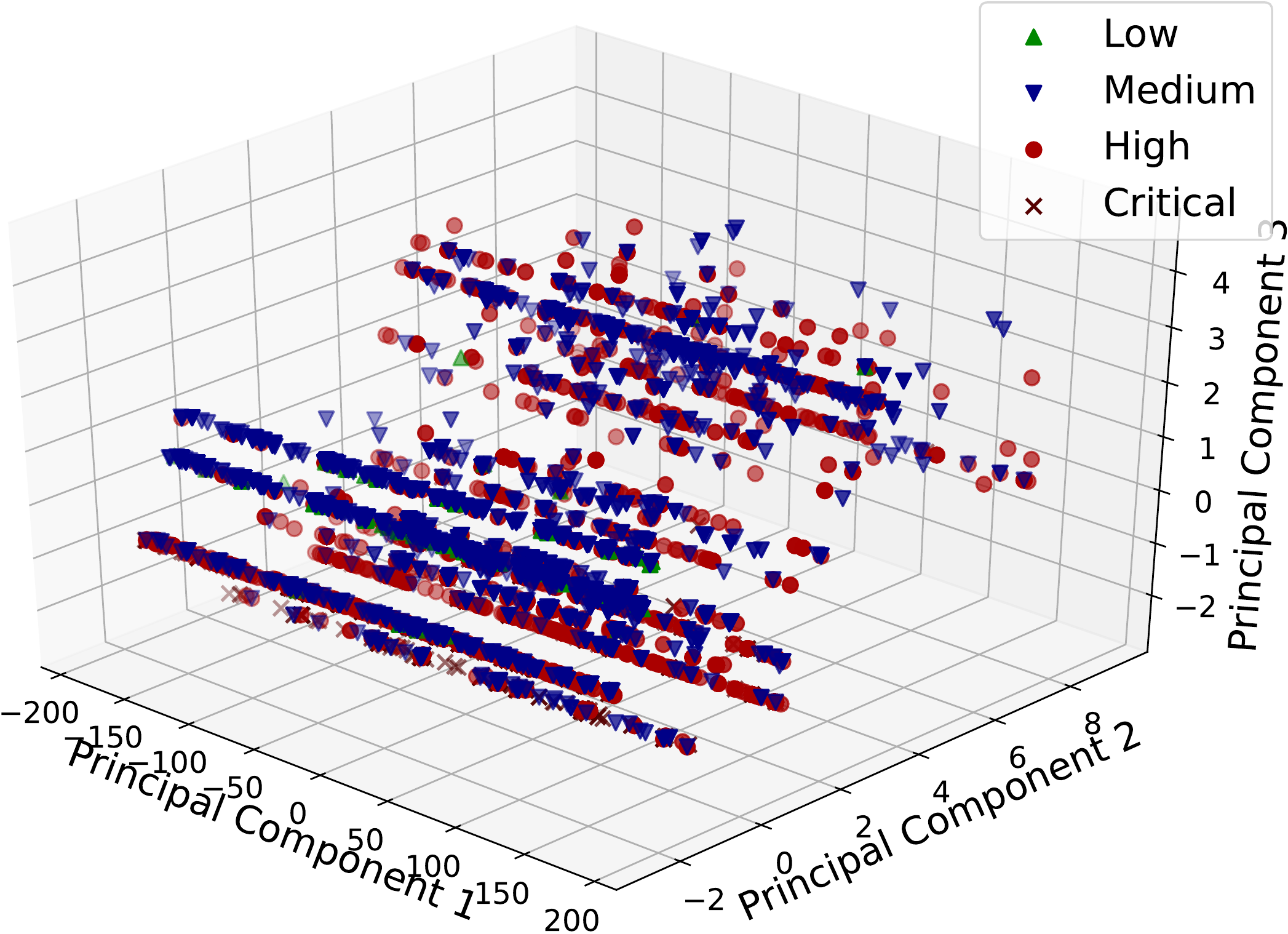}}
		\subfigure[High \label{fig:cvssMovements:High}] {\includegraphics[width=0.32\textwidth]{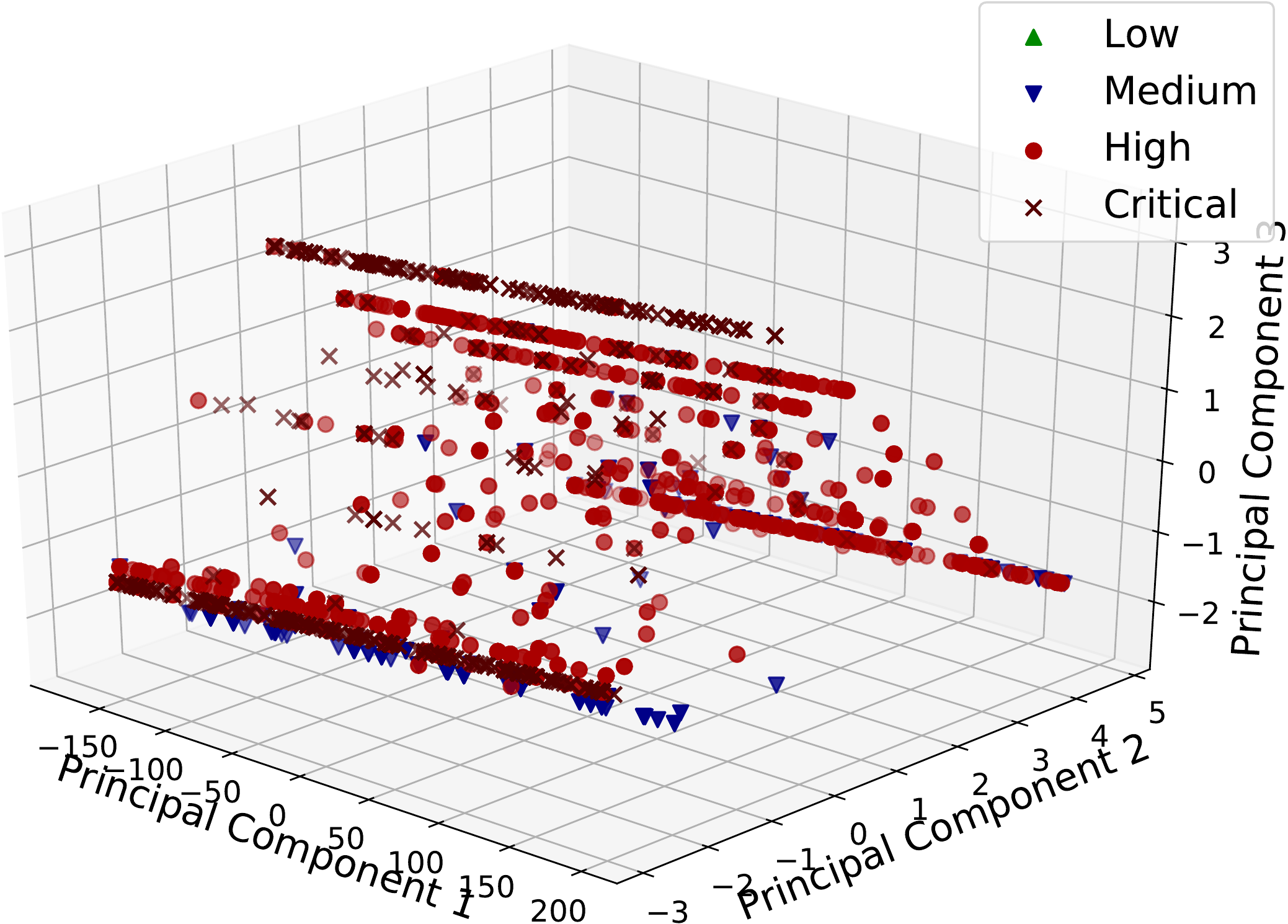}}
\caption{Vulnerabilities from Low, High, and Medium severity in CVSS v2 that transformed into different severity levels in v3. A non-linear pattern can be observed among the vulnerabilities that were assigned respective v3 severity.}
\label{fig:cvssMovements}
\end{figure*}

Vulnerabilities switch severity labels across versions due to the introduction of new parameters as well as the use of different weights to existing parameters.  
Given that v2 and v3 capture behavioral aspects of vulnerabilities, we investigated if the added parameters in v3 depend on the v2 metrics. To enrich the investigation for this extrapolation, we also used the vulnerability type information of every vulnerability. Then, we explored the patterns within a v2 label that lead to a change in severity. To visualize the patterns, we began by applying the Principal Component Analysis (PCA) as a feature reduction technique. PCA is a linear dimensionality reduction technique using the Singular Value Decomposition (SVD) of the data to project it to a lower-dimensional space~\cite{TippingB99}, reducing the 13-dimensional feature vector to a three-dimension space. \autoref{fig:cvssMovements} shows the features in a 3D space. We utilize the 3-D representation because the 2-D representation had overlapping severity labels, thereby making it difficult to distinguish the different patterns. 
For example, the figure shows the different v3 labels a High (v2) severity vulnerability has moved to. While the vulnerabilities in v2 Low are scattered in the space, High and Medium in v2 have followed specific and clear patterns. This means that vulnerabilities with Low v2 severity scores were the most affected by the v3 transformation. These patterns indicate that the added parameters in the v3 severity calculation can be extrapolated from the existing v2 parameters. Moreover, the scattered distribution of vulnerabilities with Low severity in v2 highlights the fundamental changes applied in v3.

\begin{table}[h]
\begin{center}
\caption{Ground truth - prediction results}
\label{tab:v2V3MoveOverall}
\scalebox{0.7}{
\begin{tabular}{l|rr|rr|rr|rr}
\toprule
\multirow{2}{*}{\backslashbox{v2}{v3}}  & \multicolumn{2}{c|}{L} & \multicolumn{2}{c|}{M} & \multicolumn{2}{c|}{H} & \multicolumn{2}{c}{C} \\

  & \multicolumn{1}{c}{\#}       & \multicolumn{1}{c|}{\%}         & \multicolumn{1}{c}{\#}        & \multicolumn{1}{c|}{\%}        & \multicolumn{1}{c}{\#}        & \multicolumn{1}{c|}{\%}        & \multicolumn{1}{c}{\#}        & \multicolumn{1}{c}{\%}        \\\midrule
L & 3 & 0.08  & 3823     & 98.76 & 45        &  1.16  & 0         & 0.00      \\
M & 0   & 0.00  & 9724   & 42.77  & 13010    & 57.23 & 0         & 0.00 \\
H & 0   & 0.00  & 320   &   2.87    & 5438   & 48.70  & 5409    & 48.43  \\
\bottomrule
\end{tabular}}
\end{center}
\end{table}

\subsection{Prediction Performance}\label{app:predictPerf}
In table \ref{tab:v2Movements}, we observed that the movement of v2 vulnerabilities with High severity level is $\approx$equally split between High and Critical severity levels when transformed to v3.  However, the prediction results of the vulnerabilities with no v3 severity in table \ref{tab:v2MovementsPV3} shows that the split of v2 vulnerabilities with High severity that transform to critical severity level is $\approx$twice the number of vulnerabilities that transform to High severity in v3. To ensure the performance of our prediction, we check the behavior of the model for the ground truth dataset. We begin by using our model to predict for the vulnerabilities that have v3 labeled. \autoref{tab:v2V3MoveOverall} shows the results of this experiment. Recall from \autoref{tab:v2Movements} that only 1\% of v2-medium and 9.5\% v2-low vulnerabilities transformed to low severity level in v3. We, therefore, see less number of vulnerabilities in the v3 low severity level. Considering that this experiment includes the training dataset, which makes 80\% of our overall dataset, we now look into only the testing dataset, removing possible biases. \autoref{tab:v2V3MoveTestGT} shows the actual representation of the ground truth-testing dataset, while table \ref{tab:v2V3MoveTest} shows the movements of the same vulnerabilities by our prediction model. Notice that low severity vulnerabilities in v2 are only 10\% of the total testing dataset, out of which, only 1.38\% of the samples remain in low in v3, leading to most of the low vulnerabilities in v2 moving to medium severity level in v3. Observe that in the tables, \ref{tab:v2V3MoveOverall}, and \ref{tab:v2V3MoveTest}, we see that the v2-high vulnerabilities have proportionally transformed to v3-high and v3-critical. Considering these the only explanation for the presence of $\approx$twice the number of transformed v3-critical vulnerabilities than v3-high (from v2-high) is the nature of their feature space than possible aberration in our model.

\begin{table}[t]
\begin{center}
\caption{Test dataset - ground truth data}
\label{tab:v2V3MoveTestGT}
\scalebox{0.7}{
\begin{tabular}{l|rr|rr|rr|rr}
\toprule
\multirow{2}{*}{\backslashbox{v2}{v3}}  & \multicolumn{2}{c|}{L} & \multicolumn{2}{c|}{M} & \multicolumn{2}{c|}{H} & \multicolumn{2}{c}{C} \\

  & \multicolumn{1}{c}{\#}       & \multicolumn{1}{c|}{\%}         & \multicolumn{1}{c}{\#}        & \multicolumn{1}{c|}{\%}        & \multicolumn{1}{c}{\#}        & \multicolumn{1}{c|}{\%}        & \multicolumn{1}{c}{\#}        & \multicolumn{1}{c}{\%}        \\\midrule
L & 104 & 13.42  & 644     & 83.10 &    27     & 3.48  & 0         & 0.00      \\
M & 84   & 1.85  & 2,368   & 52.08    & 1,974    & 43.41 & 121         & 2.66 \\
H & 0   & 0.00  & 85   &   3.80    & 950   & 42.52  & 1,199    & 53.67  \\
\bottomrule
\end{tabular}}
\end{center}
\end{table}

\begin{table}[h]
\begin{center}
\caption{Test dataset - prediction results}
\label{tab:v2V3MoveTest}
\scalebox{0.7}{
\begin{tabular}{l|rr|rr|rr|rr}
\Xhline{2\arrayrulewidth}
\multirow{2}{*}{\backslashbox{v2}{v3}}  & \multicolumn{2}{c|}{L} & \multicolumn{2}{c|}{M} & \multicolumn{2}{c|}{H} & \multicolumn{2}{c}{C} \\
\cline{2-9}
  & \multicolumn{1}{c}{\#}       & \multicolumn{1}{c|}{\%}         & \multicolumn{1}{c}{\#}        & \multicolumn{1}{c|}{\%}        & \multicolumn{1}{c}{\#}        & \multicolumn{1}{c|}{\%}        & \multicolumn{1}{c}{\#}        & \multicolumn{1}{c}{\%}        \\
  \Xhline{2\arrayrulewidth}

L & 6 & 0.77  & 765     & 98.71 & 4        &  0.52  & 0         & 0.00      \\
M & 0   & 0.00  & 2128   & 46.80  & 2419    & 53.20 & 0         & 0.00 \\
H & 0   & 0.00  & 58   &  2.6     & 933   &  47.76 & 1243    &  55.64 \\
\Xhline{2\arrayrulewidth}
\end{tabular}
}
\end{center}
\end{table}

\subsection{Impact: Vendor and Product Consistency}\label{app:vendorConsisCaseStudy}
Recall that in section~\ref{sec:vendorIncons}, we identify, quantify, and remedy the inconsistencies in vendor and product names in NVD. The vulnerabilities corresponding to the inconsistent vendor names are assigned to the consistent vendors (identified by vulnerability count). What type of vulnerabilities are impacted by such inconsistencies? Are they unimportant so that they can be considered as those that may not have much impact on host systems and can thus be ignored? To answer these questions, we consider the vulnerabilities that have inconsistent vendor or product names. Among those that are corresponding to well-known vendors, we select 10 CVEs randomly, shown in~\autoref{app:tab:caseStudy}. To evaluate their impact, we focus on their severity and vulnerability type. Notice that all except one (CVE-2006-6601) are of High severity (v2). This CVE-2006-6601 vulnerability is in windows media player though of Medium severity, which can be exploited by a crafted header of .MID (MIDI) file to and cause a DoS attack. Among the other nine vulnerabilities, four can be exploited remotely. Additionally, CVE-2018-16983, a vulnerability in tor browser, can be exploited by an attacker to bypass by using text/html;/json Content-Type, which can pose to be a privacy risk. 

These analyses show that the vulnerabilities corresponding to the inconsistent vendor names are impacting, severe, and thus cannot be ignored. Additionally, it exhibits the importance of having a consistent vendor/product name.

\begin{table}[h]
\centering
\caption{Case study: A sample of vulnerabilities corresponding to known vendors. These vendors were mislabelled, meaning that they have another instance of its own. For example, the dominant instance of microsft is microsoft. We uniform the dominant instance as the consistent vendor name. Most of these vulnerabilities give remote access to the adversary.}
\label{app:tab:caseStudy}
\scalebox{0.78}{
\begin{tabular}{llll}
\toprule
CVEs           & Vendor              & Severity (v2) & Description                                   \\
\midrule
CVE-2017-7689  & schneider\_electric & High     & Command injection                             \\
CVE-2006-6601  & windows             & Medium   & Malformed header (DoS)                                           \\
CVE-2008-4019  & microsft            & High     & Remote code execution                         \\
CVE-2008-3471  & microsft            & High     & Remote code execution                         \\
CVE-2014-0754  & chneider\_electric  & High     & Directory traversal \\
CVE-2009-1185  & kernel              & High     & Privilege escalation                          \\
CVE-2018-16983 & torproject          & High     & Bypass script blocking                                          \\
CVE-2008-0166  & openssl\_project    & High     & Crypto keys-based attack    \\
CVE-2017-5005  & quick\_heal         & High     & Remote code execution                         \\
CVE-2017-8774  & quick\_heal         & High     & Memory corruption   
        \\
\bottomrule
\end{tabular}}
\end{table}

\subsection{Observations: Inconsistent Vendor and Product} 
From our analysis, we observed several interesting naming patterns that reflect the complex software ecosystem and highlight difficulties that can arise in managing vendor and product names. For example: 
\begin{enumerate*}[label=\protect\circled{\arabic*}, font=\small\bfseries ]
    \item In the NVD, various entities may be deemed the vendor. Interestingly, a primary software developer is sometimes listed as a vendor, and different maintainers over time may list the same product. For example, Igor Sysoev was the original author of nginx, which is now maintained by nginx.inc, and both of them are listed as vendors with nginx as a product. Additionally, developers can be referenced with variations of their real name, leading to inconsistency (e.g., {\em provos} and {\em neilsprovos}). Acquired companies can also be listed as products under the acquiring vendor (e.g., {\em ICQ} and {\em AOL}). Note that our vendor heuristics allow us to select these vendor pairs for manual analysis.
    \item A vendor could be a parent company while the product is the subsidiary. Here, the subsidiary can be both a vendor (listing its own software) as well as a product, which is also detected by our vendor heuristics.
    \item A vendor could change name (e.g., {\em cat} became {\em quickheal}). We note that our vendor heuristics may catch this if the old and new vendor names share characters or product names, but may miss cases otherwise.
\end{enumerate*}

Thus, the NVD would benefit from defining consistent rules for vendor and product naming, such as on the use of white spaces, special characters, and abbreviations. One path forward would be to require vulnerability reporters to check their name submissions against a tool or online interface that searches existing names that likely match, perhaps using an approach such as our identification method.

\end{document}